\documentclass[longauth]{aa}

\usepackage{graphicx}
\usepackage{txfonts}
\usepackage{natbib}
\usepackage[colorlinks=true,citecolor=blue,breaklinks=true]{hyperref}
\usepackage{xspace}
\usepackage{amsmath,amssymb}
\usepackage{gensymb}
\usepackage{enumitem}
\usepackage{upgreek}

\bibpunct{(}{)}{;}{a}{}{,}

\newcommand{\MJup}{\ensuremath{M_{\mathrm{Jup}}}\xspace}
\newcommand{\Teff}{\ensuremath{T_{\mathrm{e\!f\!f}}}\xspace}
\newcommand{\mic}{\ensuremath{\upmu\mathrm{m}}\xspace}
\newcommand{\as}{\hbox{$^{\prime\prime}$}\xspace}
\newcommand{\loD}{\hbox{$\lambda/D$}\xspace}
\newcommand{\degre}{\degree\xspace}
\newcommand{\degreC}{\degree C\xspace}
\newcommand{\microrad}{\ensuremath{\upmu\mathrm{rad}}\xspace}
\newcommand{\kms}{\ensuremath{\mathrm{km}\,\mathrm{s}^{-1}}\xspace}
\newcommand{\crires}{CRIRES\xspace}

\begin{document}

\title{First light of VLT/HiRISE: \\ High-resolution spectroscopy of young giant exoplanets}
\titlerunning{First light of VLT/HiRISE}

\author{
  A.~Vigan\inst{\ref{lam}}
  \and
  M.~El Morsy\inst{\ref{lam}}
  \and
  M.~Lopez\inst{\ref{lam}}
  \and
  G.~P.~P.~L.~Otten\inst{\ref{taiwan},\ref{lam}}
  \and 
  J.~Garcia\inst{\ref{lam}}
  \and
  J.~Costes\inst{\ref{lam}}
  \and
  E.~Muslimov\inst{\ref{oxf},\ref{lam},\ref{kazan}}
  \and
  A.~Viret\inst{\ref{lam}}
  \and
  Y.~Charles\inst{\ref{lam}}
  \and
  G.~Zins\inst{\ref{esog}}
  \and
  G.~Murray\inst{\ref{durham}}
  \and
  A.~Costille\inst{\ref{lam}}
  \and
  J.~Paufique\inst{\ref{esog}}
  \and
  U.~Seemann\inst{\ref{esog}}
  \and
  M.~Houllé\inst{\ref{lam}}
  \and
  H.~Anwand-Heerwart\inst{\ref{gott}}
  \and
  M.~Phillips\inst{\ref{exeter}, \ref{ifa}}
  \and
  A.~Abinanti\inst{\ref{lam}}
  \and
  P.~Balard\inst{\ref{lam}}
  \and
  I.~Baraffe\inst{\ref{exeter},\ref{ens}}
  \and
  J.-A.~Benedetti\inst{\ref{lam}}
  \and
  P.~Blanchard\inst{\ref{lam}}
  \and
  L.~Blanco\inst{\ref{esoc}}
  \and
  J.-L.~Beuzit\inst{\ref{lam}}
  \and
  E.~Choquet\inst{\ref{lam}}
  \and
  P.~Cristofari\inst{\ref{lam}}
  \and
  S.~Desidera\inst{\ref{padova}}
  \and
  K.~Dohlen\inst{\ref{lam}}
  \and
  R.~Dorn\inst{\ref{esog}}
  \and
  T.~Ely\inst{\ref{lam}}
  \and
  E.~Fuenteseca\inst{\ref{esoc}}
  \and
  N.~Garcia\inst{\ref{lam}}
  \and
  M.~Jaquet\inst{\ref{lam}}
  \and
  F.~Jaubert\inst{\ref{lam}}
  \and
  M.~Kasper\inst{\ref{esog}}
  \and
  J.~Le~Merrer\inst{\ref{lam}}
  \and
  A.-L.~Maire\inst{\ref{ipag}}
  \and
  M.~N'Diaye\inst{\ref{oca}}
  \and
  L.~Pallanca\inst{\ref{esoc}}
  \and
  D.~Popovic\inst{\ref{esog}}
  \and
  R.~Pourcelot\inst{\ref{stsci},\ref{lam}}
  \and
  A.~Reiners\inst{\ref{gott}}
  \and
  S.~Rochat\inst{\ref{ipag}}
  \and
  C.~Sehim\inst{\ref{lam}}
  \and
  R.~Schmutzer\inst{\ref{esoc}}
  \and
  A.~Smette\inst{\ref{esoc}}
  \and
  N.~Tchoubaklian\inst{\ref{lam}}
  \and
  P.~Tomlinson\inst{\ref{lam}}
  \and
  J.~Valenzuela~Soto\inst{\ref{esoc}}
}

\institute{
  Aix Marseille Univ, CNRS, CNES, LAM, Marseille, France \label{lam}
  \\ \email{\href{mailto:arthur.vigan@lam.fr}{arthur.vigan@lam.fr}}
  \and
  Institute for Astrophysics, Georg-August University, Friedrich-Hund-Platz 1, 37077 Göttingen, Germany \label{gott}
  \and
  European Southern Observatory (ESO), Karl-Schwarzschild-Str. 2, 85748 Garching, Germany \label{esog}
  \and
  Center for Advanced Instrumentation, Durham University, Durham, DH1 3LE, United Kindgom \label{durham}
  \and
  European Southern Observatory, Alonso de Cordova 3107, Vitacura, Santiago, Chile \label{esoc}
  \and
  Physics \& Astronomy Dpt, University of Exeter, Exeter, EX4 4QL, UK \label{exeter}
  \and
  Universit\'e C\^ote d'Azur, Observatoire de la C\^ote d'Azur, CNRS, Laboratoire Lagrange, France \label{oca}
  \and
  Univ. Grenoble Alpes, CNRS, IPAG, F-38000 Grenoble, France \label{ipag}
  \and
  INAF -- Osservatorio Astronomico di Padova, Vicolo dell'Osservatorio 5, 35122 Padova, Italy \label{padova}
  \and
  Academia Sinica, Institute of Astronomy and Astrophysics, 11F Astronomy-Mathematics Building, NTU/AS campus, No. 1, Section 4, Roosevelt Rd., Taipei 10617, Taiwan \label{taiwan}
  \and
  Dept. of Astrophysics, University of Oxford, Keble Road, Oxford, OX1 3RH, UK \label{oxf}
  \and
  Optical and Electronic Systems Department, Kazan National Research Technical University \label{kazan}
  \and
  Institute for Astronomy, University of Hawaii at Manoa, Honolulu, HI 96822, USA \label{ifa}
  \and
  École Normale Supérieure, Lyon, CRAL (UMR CNRS 5574), Université de Lyon, France \label{ens}
  \and
  Space Telescope Science Institute, 3700 San Martin Drive, Baltimore, MD 21218, USA \label{stsci}
}
\date{Received 19 September 2023; accepted 24 October 2023}

\abstract{
  A major endeavor of this decade is the direct characterization of young giant exoplanets at high spectral resolution to determine the composition of their atmosphere and infer their formation processes and evolution. Such a goal represents a major challenge owing to their small angular separation and luminosity contrast with respect to their parent stars. Instead of designing and implementing completely new facilities, it has been proposed to leverage the capabilities of existing instruments that offer either high-contrast imaging or high-dispersion spectroscopy by coupling them using optical fibers. In this work, we present the implementation and first on-sky results of the High-Resolution Imaging and Spectroscopy of Exoplanets (HiRISE) instrument at the Very Large Telescope (VLT), which combines the exoplanet imager SPHERE with the recently upgraded high-resolution spectrograph \crires using single-mode fibers. The goal of HiRISE is to enable the characterization of known companions in the $H$ band at a spectral resolution on the order of $R = \lambda/\Delta\lambda = 100\,000$ in a few hours of observing time. We present the main design choices and the technical implementation of the system, which is constituted of three major parts: the fiber injection module inside of SPHERE, the fiber bundle around the telescope, and the fiber extraction module at the entrance of \crires. We also detail the specific calibrations required for HiRISE and the operations of the instrument for science observations. Finally, we detail the performance of the system in terms of astrometry, temporal stability, optical aberrations, and transmission, for which we report a peak value of $\sim$3.9\% based on sky measurements in median observing conditions. Finally, we report on the first astrophysical detection of HiRISE to illustrate its potential.
}

\keywords{
  Instrumentation: high angular resolution --
  Instrumentation: adaptive optics --
  Instrumentation: spectrographs --
  Techniques: high-angular resolution --
  Techniques: spectroscopy --
  Infrared: planetary systems
}

\maketitle

\section{Introduction}
\label{sec:introduction}

Exoplanets are ubiquitous in our galaxy \citep{Mayor2011,Cassan2012,Bryson2021}, but the processes that govern their formation, drive their evolution, and could ultimately favor the apparition of life are still extremely uncertain. Understanding the origin and physics of extrasolar giant planets (EGPs) is crucial as they gravitationally dominate their systems, which can either enhance or inhibit the subsequent formation of telluric planets with masses and separations comparable to those of the Earth \citep[e.g.,][]{Walsh2011}.

Over the past two decades, multiple imaging surveys using large ground-based telescopes equipped with adaptive optics systems have uncovered a wealth of substellar companions that range from a few tens down to a few Jupiter masses \citep{Bowler2016,Chauvin2018,Hinkley2021}. These surveys have also enabled the constraining of their formation pathway from the statistical point of view \citep{Nielsen2019,Vigan2021}. However, the number of detections remains small, even with the most recent instruments such as VLT/SPHERE \citep{Beuzit2019}, Gemini/GPI \citep{Macintosh2014}, or Subaru/SCExAO \citep{Jovanovic2015}, which prevents us from drawing strong conclusions at a population-level based on statistical inference.

A complementary approach to understanding EGPs is to study their atmospheres, which bear unique markers of their formation mechanisms \citep[e.g.,][]{Oberg2011}, internal structure \citep[e.g.,][]{Madhusudhan2019}, bulk properties \citep[e.g.,][]{Burrows1997}, or ongoing chemical and dynamical processes \citep[e.g.,][]{Moses2011,Phillips2020}. Some of these markers can be detected and quantified at relatively low spectral resolution ($R = \Delta\lambda / \lambda \leq 100$), which is readily accessible in current high-contrast instruments equipped with integral-field spectrographs (IFS). However, spectral resolutions higher than a few thousand are necessary for more detailed measurements: detection of isotopologues \citep{Molliere2019}, determination of abundances \citep{Konopacky2013}, measurement of radial velocities to study orbital elements \citep{Ruffio2019}, looking for exo-moons \citep{Vanderburg2018,Lazzoni2022,Ruffio2023}, or determination of rotational velocities \citep{Bryan2018}.

High-dispersion spectroscopy (HDS) for the detection of exoplanetary signal was suggested more than two decades ago as a way to boost the detection capabilities by relying on the numerous spectral features expected in the spectra of exoplanets \citep{Sparks2002,Riaud2007}. This method was first applied to detect the thermal irradiation of transiting and non-transiting close-in EGPs \citep{Snellen2010,Brogi2012,Birkby2013}, before being used to characterize the direct near-infrared (NIR) emission of young EGPs detected with high-contrast imaging \citep{Snellen2014,Schwarz2016}. This method was then generalized and applied to integral-field-unit data to boost the detection capabilities of existing \citep[e.g.,][]{Hoeijmakers2018} and forthcoming \citep[e.g.,][]{Houlle2021} instruments.

To combine the potential of high-contrast imaging (HCI) with HDS for EGPs study, several projects have proposed coupling existing adaptive optics (AO) instruments with high-resolution spectrographs using single-mode optical fibers (SMF). The KPIC instrument on Keck \citep{Delorme2021} has been in operation for a few years and has already provided several astrophysical results in the $K$ band \citep[e.g.,][]{Wang2021}. Similarly, the REACH instrument on Subaru is also operational \citep{Kotani2020} and providing observations in the $H$ band.

The unit 3 (UT3 -- Melipal) of the Very Large Telescope (VLT) hosts two instruments ideally suited for this purpose. On one side, the Spectro-Polarimetric High-contrast Exoplanet REsearch instrument \citep[SPHERE;][]{Beuzit2019} is attached to the Nasmyth A focus. On the other side, following the CRIRES$^{+}$ project \citep{Kaeufl2004,Dorn2014,Dorn2023}, the upgraded CRyogenic high-resolution Infra-Red Spectrograph (CRIRES) was attached to the Nasmyth B focus. HiRISE implements of a fiber coupling between these two instruments to enable the characterization of known companions at a spectral resolution of $R \simeq 100\,000$.

In this paper, we present the HiRISE instrument and its performance after commissioning on the VLT. The main astrophysical requirements and design choices for the system are described in Sect.~\ref{sec:design_choices}. Then, in Sect.~\ref{sec:implementation} we present the implementation of the system, and in Sect.~\ref{sec:operations} we state how the system is calibrated and operated. The on-sky performance and first astrophysical results are detailed in Sect~\ref{sec:performance}. Finally, we conclude and present some perspectives in Sect.~\ref{sec:conclusions}.

\section{Design choices}
\label{sec:design_choices}

The idea behind HiRISE is to benefit from the HCI capabilities of SPHERE and the HDS capabilities of \crires to perform the spectral characterization of known EGPs at high spectral resolution. To guide the overall design of HiRISE, we derived three astrophysical top-level requirements (TLRs):

\begin{enumerate}[leftmargin=*,label={\texttt{sci.req.\arabic*}}]
  \item The instrument must enable the direct characterization of substellar companions, at a significance level higher than 5$\sigma$, with integration times shorter than the duration of one night (typically 8 hours). \label{SR1}
  \item The instrument must be more efficient than \crires in standalone mode for the same science case. \label{SR2}
  \item The instrument must provide access to the $H$ band and, if possible, to the $K$ band. \label{SR3}
\end{enumerate}

These requirements were explored in detail by \citet{Otten2021} based on a preliminary design of HiRISE. They demonstrated, in particular, that the overall transmission of the system is a driving parameter of the final performance. Their analysis showed that \ref{SR1} is easily satisfied for bright stars ($H < 7$) and faint companions; typical exposure times of two to three hours are sufficient to detect companions even at very high contrast, such as 51\,Eri\,b \citep{Macintosh2015}. They also showed that in the $H$ band, there is a clear part of the contrast versus angular separation parameter space where HiRISE significantly outperforms \crires in standalone mode, therefore satisfying \ref{SR2}. In the $K$ band, the noise budget is dominated by the thermal emission of the sky, the SPHERE instrument, and the output face of the science fibers \citep[][their Sect.~2.8 and Fig.~7]{Otten2021}.

The $K$ band is also constraining from the point of view of optical fibers. Classical telecom fibers made of extremely pure fused silica are highly transmissive below 1.7--1.8\,\mic. Beyond that, their attenuation quickly increases, making them unsuitable for the $K$ band and lengths longer than a few meters. Fluoride glass fibers (often referred to as ZBLAN fibers) are optimized for longer wavelengths, but usually come at a high price and have a reputation of being difficult to work with, driving up the costs and manufacturing risks. Based on discussions with the French manufacturer \emph{\emph{Le Verre Fluoré}}, the preliminary design of HiRISE considered using ZBLAN fibers that would enable observations in both the$H$ and $K$ bands. However, \citet{Otten2021} demonstrated that even with such fibers, \crires in standalone mode would remain more efficient than HiRISE in the $K$ band, and \ref{SR2} would therefore not be satisfied.

From the astrophysical point of view, the $K$ band is particularly interesting due to the prominent CO features starting at around 2.3\,\mic. Other species, such as H$_2$O or CH$_4,$ also have signatures in the $K$ band, but they are much weaker and dominated by the CO. On the contrary, C- and O-bearing species have more balanced absorption features in the $H$ band, which makes this wavelength range interesting to study. Moreover, future detections coming from the combination of radial velocity and ESA/Gaia astrometry \citep[e.g.,][]{Rickman2022} will probably be older and colder than currently known EGPs, possibly exhibiting bluer colors with a peak emission in the $H$ or even $J$ band. This makes the $H$ band interesting for HiRISE in the long term.

Taking into account the lower expected performance level and the additional costs, risks, and possible delays associated with an implementation of the $K$ band in HiRISE, the project authors decided to abandon the $K$ band and to fully optimize the design for the $H$ band. The main optimization was to use highly transmissive telecom fibers, but improvements were also made on the design of anti-reflection (AR) coatings and a dichroic filter. A more in-depth study could have been done to further optimize the $K$ -band design, but the project did not have sufficient resources at the time. In the final design, \ref{SR3} is therefore only partly satisfied.

The HiRISE instrument uses single-mode fibers (SMFs) instead of multimode fibers (MMF). This choice was driven by the spatial filtering properties of SMFs \citep[e.g.,][]{CoudeduForesto1994,Patru2008}. In the context of HCI, \citet{Mawet2017} demonstrated the significant gain in contrast that is obtained ``for free'' due to the poor projection of residual atmospheric phase onto the mode of the fiber compared to the projection of a well-centered planetary point-spread function \citep{Jovanovic2015}. The use of SMF also significantly reduces the impact of modal noise, focal ratio degradation, and transmission of thermal background. Moreover, SMFs in spectroscopy produce a well-defined, quasi-Gaussian, and very stable line-spread function, which offers many advantages for spectral extraction and Doppler shift measurements. However, SMFs induce strong constraints in terms of centering of the planetary PSF on the fiber, which requires using carefully calibrated centering strategies \citep{ElMorsy2022}.

The very small core of SMF, typically 4--8\,\mic for telecom fibers, also implies that connectors should be avoided to reduce flux losses. For a fiber with a 6~\mic core, a decentering of only 0.5\,\mic between two connected fibers will induce a loss on the order of 10\% in the transmitted flux (0.46\,dB). This is without even considering Fresnel losses at the interface between the two fibers. In HiRISE, although connectors were initially considered to simplify the installation of the fiber bundle, they were eventually dropped due to the technical challenge of reducing losses to less than 0.1\,dB, or even 0.2\,dB.

\begin{figure*}
  \centering 
  \includegraphics[height=8.9cm]{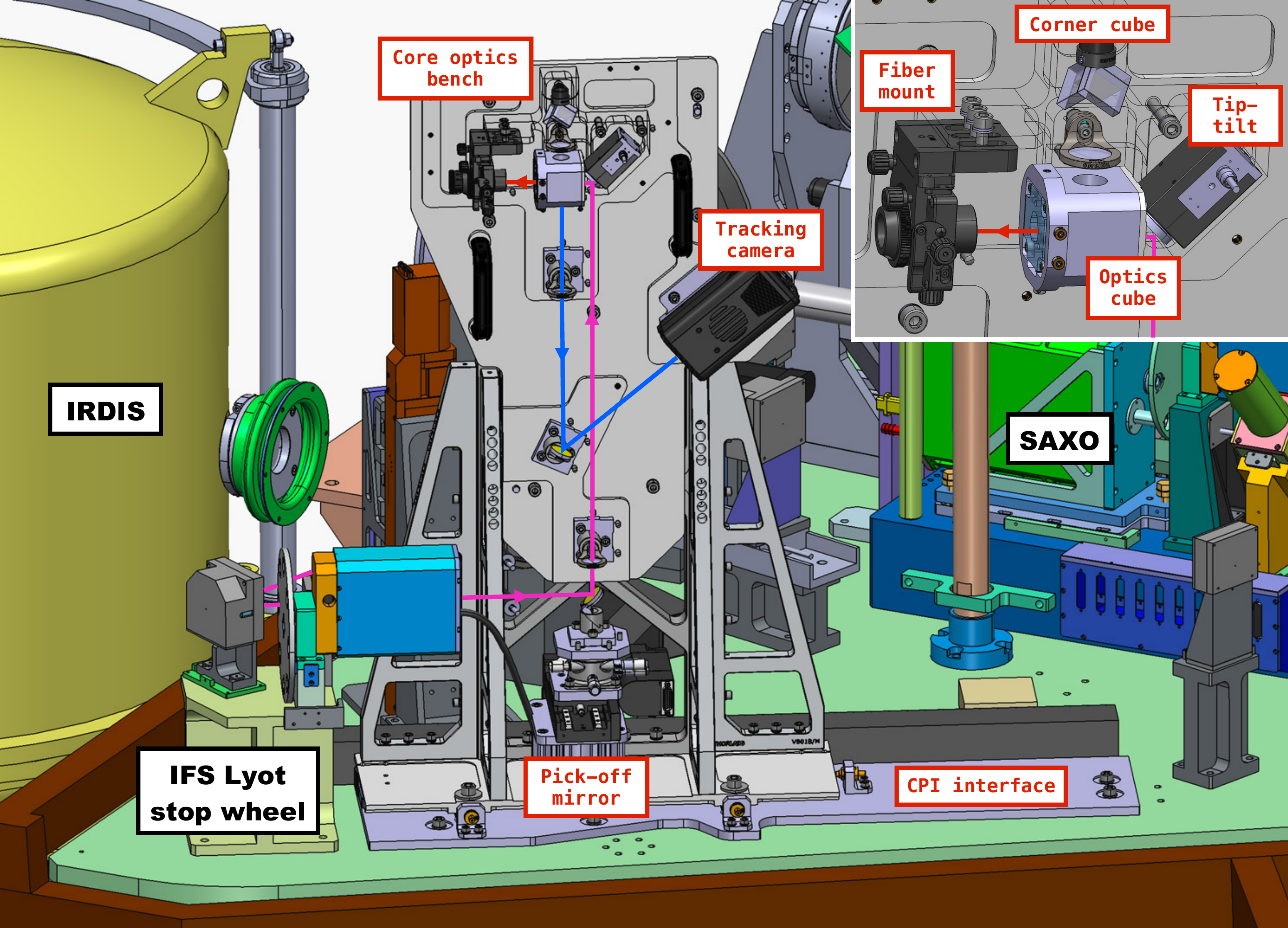}
  \includegraphics[height=8.9cm]{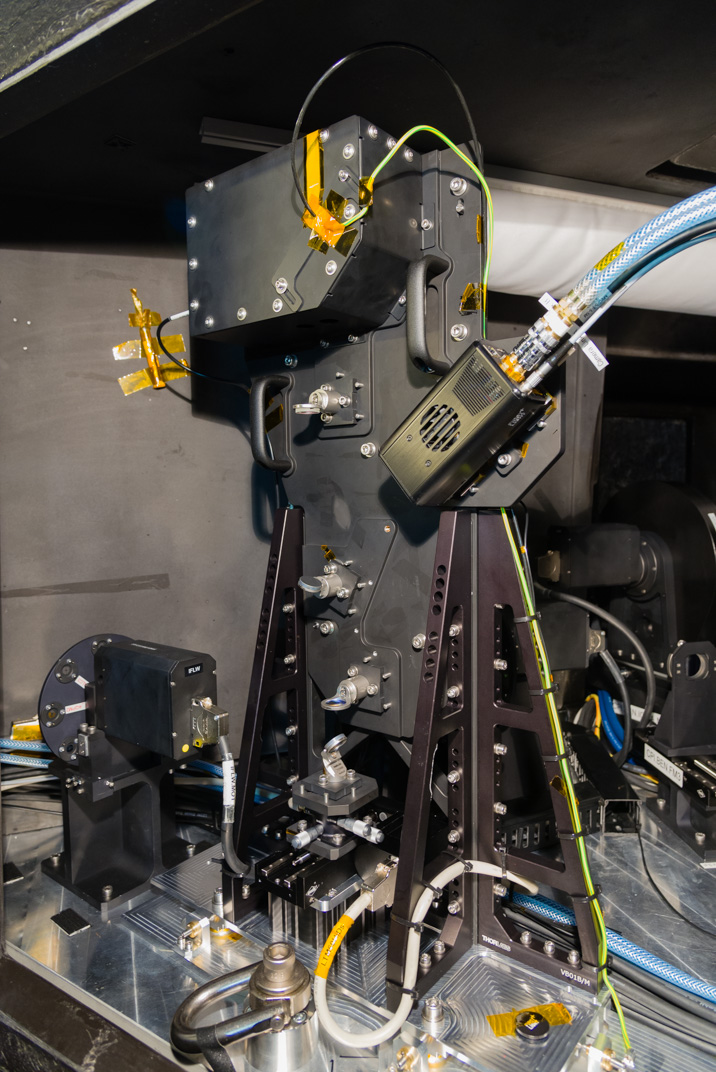}
  \caption{Implementation of the FIM inside the IFS arm of SPHERE. On the left drawing, the SPHERE components are labeled in black, while the FIM components are labeled in red. The inset in the top right shows the components of the core optics bench. The light path is illustrated on the drawing in purple for $YJH$, blue for $YJ$, and red for $H$ \citep[see also][their Fig. 3]{Vigan2022spie}. The picture on the right shows the FIM at the end of the installation in SPHERE. The IFS Lyot wheel is visible in the bottom left of the picture.}
  \label{fig:fim}
\end{figure*}

\section{System implementation}
\label{sec:implementation}

At high levels, the system is composed of three distinct parts: (i)~the fiber injection module (FIM) implemented in SPHERE (Sect.~\ref{sec:fim}), (ii)~the fiber bundle (FB) that links SPHERE and \crires around the UT3 (Sect.~\ref{sec:fb}), and (iii)~the fiber extraction module (FEM) installed in \crires (Sect.~\ref{sec:fem}). However, both SPHERE and \crires are standalone VLT instruments that are operated independently. They have their own designs, constraints, and operational models. From these considerations, we derived three technical TLRs that drove the technical design of HiRISE:

\begin{enumerate}[leftmargin=*,label={\texttt{tech.req.\arabic*}}]
  \item The instrument must not impact regular operations of SPHERE, \crires, or VLT-UT3 when it is not used. \label{TR1}
  \item The instrument must not require any modification of the hardware used in regular SPHERE and \crires operations. \label{TR2}
  \item The instrument must be compatible, as much as possible, with ESO and VLT standards. \label{TR3}
\end{enumerate}

The following subsections describe the different parts of the system and how the above technical TLRs were satisfied for each of them. We describe first the FIM and its electronics cabinet, then the FB, and finally the FEM.

\subsection{Fiber injection module}
\label{sec:fim}

\begin{figure*}
  \centering 
  \includegraphics[width=0.7\textwidth]{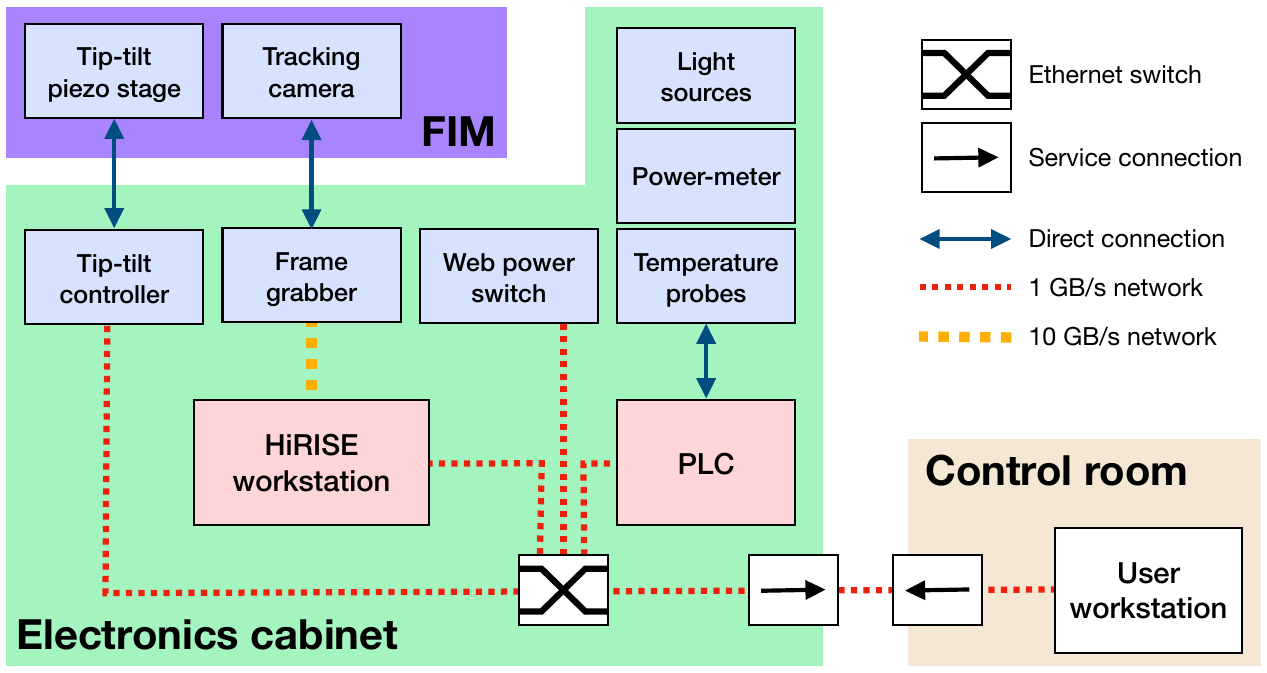}
  \caption{Connection mapping between electronics cabinet, different devices, and control network used to operate the instrument from the control room.}
  \label{fig:electronics_cabinet}
\end{figure*}

\subsubsection{Opto-mechanical implementation}

The FIM is implemented inside the SPHERE instrument, more specifically in the NIR arm that feeds the IFS. The opto-mechanical design of the FIM is illustrated in Fig.~\ref{fig:fim}, where the main components are highlighted. A more detailed view of the optical design for the FIM is available in \citet[][their Fig.~1]{ElMorsy2022}.

For the opto-mechanical design of the FIM, one of the main aspects was to strictly satisfy \ref{TR1} and \ref{TR2}. Since there was no space available upstream of the dichroic filter that separates the NIR beams between IRDIS and IFS, the FIM is located in the IFS arm, just after the IFS Lyot stop wheel where a flat space is available on the common path interface (CPI) bench. This space was, however, not sufficient to accommodate the optical design without requiring several folding mirrors, which is why the FIM bench was implemented vertically over a dedicated interface plate. The IFS beam, which has a diameter of less than 10\,mm, goes through the brackets that support the FIM bench without being vignetted. The CPI interface plate is attached to the CPI with five screws using existing threaded holes that were originally used for SPHERE alignment tools. To satisfy \ref{TR3} for SPHERE and minimize stray light reflections, mechanical parts close to optical beams were anodized with an inorganic material, which is black both in the visible and in the near-infrared.

When using HiRISE, a pick-off mirror is inserted into the IFS beam with a precision linear stage. Because the stage inserts an optical element that interacts with the SPHERE beam, the stage is connected to the existing SPHERE RMC controllers and controlled through the SPHERE control software. To maximize compatibility with old RMC controllers and \ref{TR3}, the selected stage is identical to existing hardware in SPHERE.

After the pick-off mirror, a single lens is then used to reimage the pupil of SPHERE onto a flat mirror glued on a piezo tip-tilt platform\footnote{An S-335 piezo tip-tilt platform from Physik Instrumente (PI), with the E-727 controller.}. The beam is then reflected toward the ``optics cube'', which contains a recollimating doublet, a dichroic plate inclined at 45\degre, and the air-gapped injection doublet that focuses the beam on the focal plane where the FB is located (science channel). The dichroic plate is a custom element manufactured by the Fresnel Institute in Marseille (France), which includes a dichroic filter on the first face and an AR coating on the second face. It reflects more than 80\% of the light from 0.95 to 1.25\,\mic toward the tracking channel, and transmits more than 98\% of the light from 1.45 to 1.85\,\mic towards the science channel. All the other optics in the cube are custom lenses manufactured by \emph{\emph{Optiques Fichou}} (France) and include AR coatings optimized for the $H$ band. Overall, the FIM transmits close to 90\% of the incoming light in the science channel (see Sect.~\ref{sec:transmission} for a detailed transmission budget).

The tracking channel includes an off-the-shelf lens and mirror to produce a focal plane at F/$\sim$30. The tracking camera is a C-RED\,2 manufactured by first light imaging (FLI), which is based on a Snake InGaAs detector \citep{Feautrier2017}. The camera is cooled using dedicated cooling lines that were added as part of the HiRISE installation. This makes it possible to cool the detector down to $-40$\degreC without the need to use any fan, therefore avoiding any vibrations on the SPHERE bench. With 15\,\mic pixels, the tracking camera is Nyquist-sampled at 0.99\,\mic, just slightly above the start of our working spectral range at $0.95\,\mic$.

A third channel, the feedback channel, is also implemented. It is used to retro-inject light from dedicated feedback fibers in the FB into the FIM and project their image over the tracking image on the camera. The beam goes backwards through the injection doublet, reflects on the dichroic plate, reflects in a corner cube, and is finally transmitted again through the dichroic plate before joining the tracking channel. A compensator lens was added just before the corner cube to compensate for differences in the longitudinal chromatism of the feedback and tracking channels. The wavelength of the calibration source has been selected to be in the transition region of the dichroic function at 1.3\,\mic ; since the feedback signal sees the dichroic function twice, it is important to select a wavelength where the signal will not be significantly attenuated. At 1.3\,\mic, approximately 10\% of the feedback signal is transmitted toward the tracking camera (for more details on the purpose of the feedback fibers, see Sect.~\ref{sec:operations} and \citet{ElMorsy2022}).

The optical quality of the FIM was estimated using a commercial NIR wavefront sensor from Phasics S.A. A total of $\sim$50\,nm\,rms was measured on-axis after alignment of the system on a ``SPHERE simulator'' reproducing the SPHERE IFS beam in the laboratory. The error budget is mainly dominated by a combination of astigmatism and spherical aberrations that could not be minimized during assembly, integration, and testing (AIT) in Europe. Off-axis, the wavefront error increases up to 80--100\,nm\,rms at the edge of the useful FIM field of view (FoV) of $\sim$1.5\as. Beyond 1.5\as, \citet{Otten2021} demonstrated that \crires in standalone mode is likely more efficient than HiRISE for the same science cases. After installation on SPHERE, a new measurement was performed with the same wavefront sensor, and a value of $\sim$70\,nm\,rms was measured on-axis, this time including the contribution of the SPHERE CPI optics. This is compatible with previous results reporting between 50 and 60\,nm\,rms of aberrations in the SPHERE instrument \citep{N'Diaye2016,Vigan2019,Vigan2022}.

Conceptually, the FB remains static with respect to the FIM, and it is the science image that is moved with respect to the fibers using the tip-tilt mirror. The PI S-335 stage offers a tip-tilt angle of $\pm$17.5\,mrad ($\pm$1\degre) with a resolution in closed-loop of 1\,\microrad and a linearity of 0.05\%. During AIT and commissioning at the VLT, we measured a conversion factor of $\sim$42.2\,pix/mrad between the tip-tilt mirror and the tracking camera. At F/30 and with pixels of 15\,\mic on the tracking camera, we therefore have the equivalence of $1\,\lambda/D = 3.22\,\mathrm{pix} = 76\,\upmu\mathrm{rad}$ at $\lambda = 1.6$\,\mic. The work of \citet{ElMorsy2022} showed that the specification on the centering of the PSF on the science fiber should be better than 0.1\,\loD to reach the best performance, which corresponds to 7.6\,\microrad. This is well within the accuracy of 1\,\microrad that is typically offered by the tip-tilt platform.

We highlight here the fact that SPHERE includes a set of high-quality NIR atmospheric dispersion compensators (ADCs), which offer residuals smaller than 1\,mas\,rms for zenith angles up to 60\degre (K. Dohlen, private communication). This is a critical parameter for the FIM, which requires a very accurate centering of the PSF on the science fiber. For a science target at a zenith angle of 30\degre, the differential atmospheric dispersion over the complete $H$ band is typically on the order of 20\,mas, which corresponds to almost 0.5\,\loD at the center of the band. This is well above our 0.1\,\loD specification for the centering and would therefore have a major negative impact on the performance without the ADCs.

\subsubsection{Electronics cabinet}

The FIM includes several active components that need to be controlled to operate the instrument. For this purpose, a new dedicated electronics cabinet has been implemented on the SPHERE Nasmyth platform below the SPHERE enclosure near IRDIS. This location has the advantage of leaving space for maintenance (\ref{TR1}) and of being exactly underneath the cable feed-through that is closest to the FIM. 

The cabinet embeds all the elements necessary to control, operate, and calibrate the FIM and the FEM (Fig.~\ref{fig:electronics_cabinet}). One of the key elements of the cabinet is the HiRISE workstation running Linux and the 2020 release of the VLT Software. The workstation is a fanless industrial computer used for embedded applications. It is connected to the VLT control network through a Cisco switch that is also embedded in the cabinet.

For the operations and calibrations, the cabinet includes a broadband halogen light source used by the FEM (see Sect.~\ref{sec:fem}), a LED light source at 1.3\,\mic used for the feedback fibers, a power-meter coupled with an InGaAs photodiode (see Sect.~\ref{sec:operations}), two Pt100 temperature sensors (one in the cabinet and one on the core optics bench of the FIM), the controller for the piezo tip-tilt stage, and a Camera Link-to-GigE frame grabber\footnote{The iPort CL-Ten from Pleora Technologies.} to acquire frames from the tracking camera.
The iPort CL-Ten frame grabber was selected as it is supported in the VLT Software based on developments done for the NIRPS instrument \citep{Bouchy2017,Wildi2017}. It converts the Camera Link signal into the GigE Vision standard, which is transmitted directly to the HiRISE workstation over a 10\,GB Ethernet optical link connected to a dedicated network interface.

The light sources, temperature probes, and power meter are connected to and controlled by a Beckhoff CX2100 programmable logic controller (PLC) that follows the most recent ESO standards for VLT instrumentation. The PLC is connected to the workstation through the HiRISE network switch.
The piezo tip-tilt stage controller is also connected to the network switch and is controlled over TCP/IP using the dedicated general command set (GCS) from PI.

Finally, the cabinet also has multiple power sockets, including three that are controlled through a web power switch, which is connected to the network switch and can be activated from the workstation. The tracking camera and frame grabber are both connected to this controllable power socket to be able to remotely perform power cycles and to power them down when the instrument is not in use.

\subsection{Fiber bundle}
\label{sec:fb}

\begin{figure*}
  \centering 
  \includegraphics[width=1\textwidth]{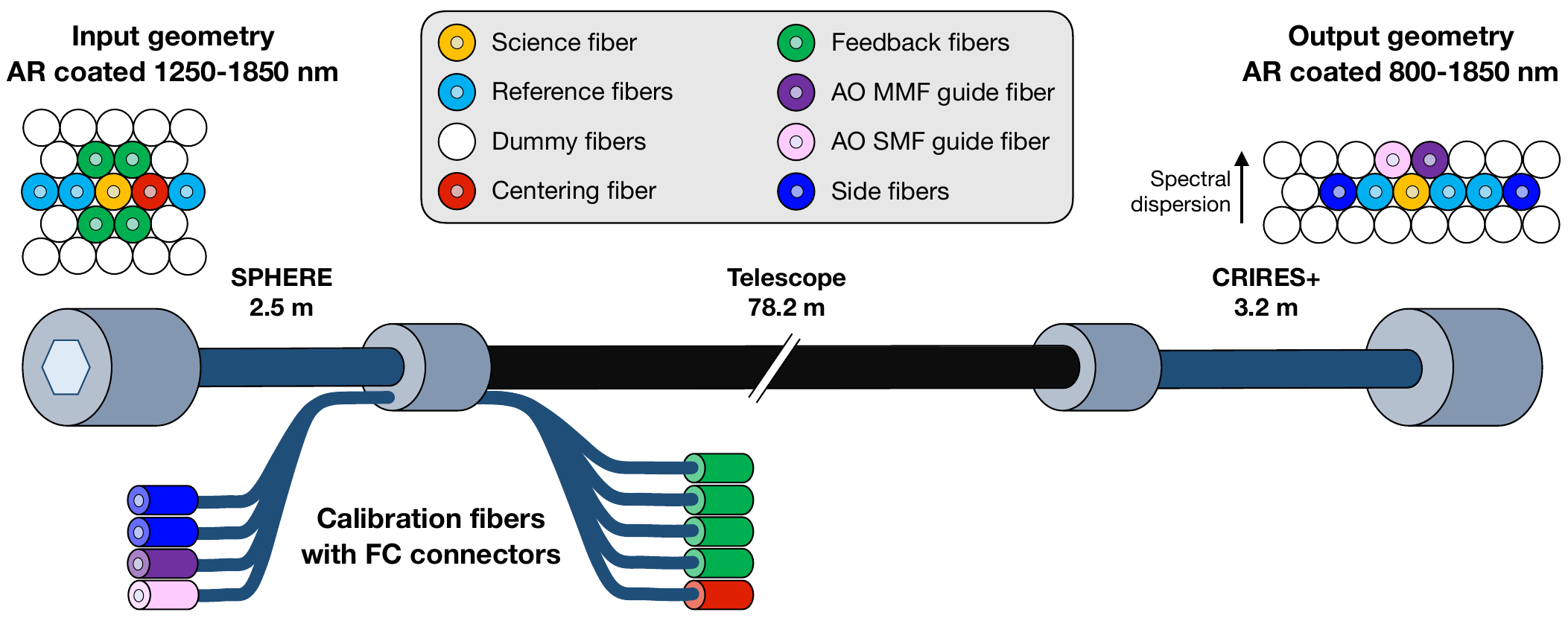}
  \caption{Conceptual drawing of the FB. Only four fibers (science and references) go all the way from the input ferrule to the output ferrule. The other fibers are for calibration or AIT, or simply dummy fibers filling the geometrical patterns in the ferrules. All calibration fibers enter or exit the bundle at the level of the junction between the SPHERE and telescope sections. The SPHERE and \crires sections have a smaller diameter (4.8\,mm) and are more flexible than the central telescope section (9.2\,mm), which is reinforced. All fibers are identical Nufern 1310M-HP SMF, except for the AO MMF guide fiber that is multimode with a 50\,\mic core.}
  \label{fig:fb_drawing}
\end{figure*}

\begin{figure*}
  \centering 
  \includegraphics[width=0.33\textwidth]{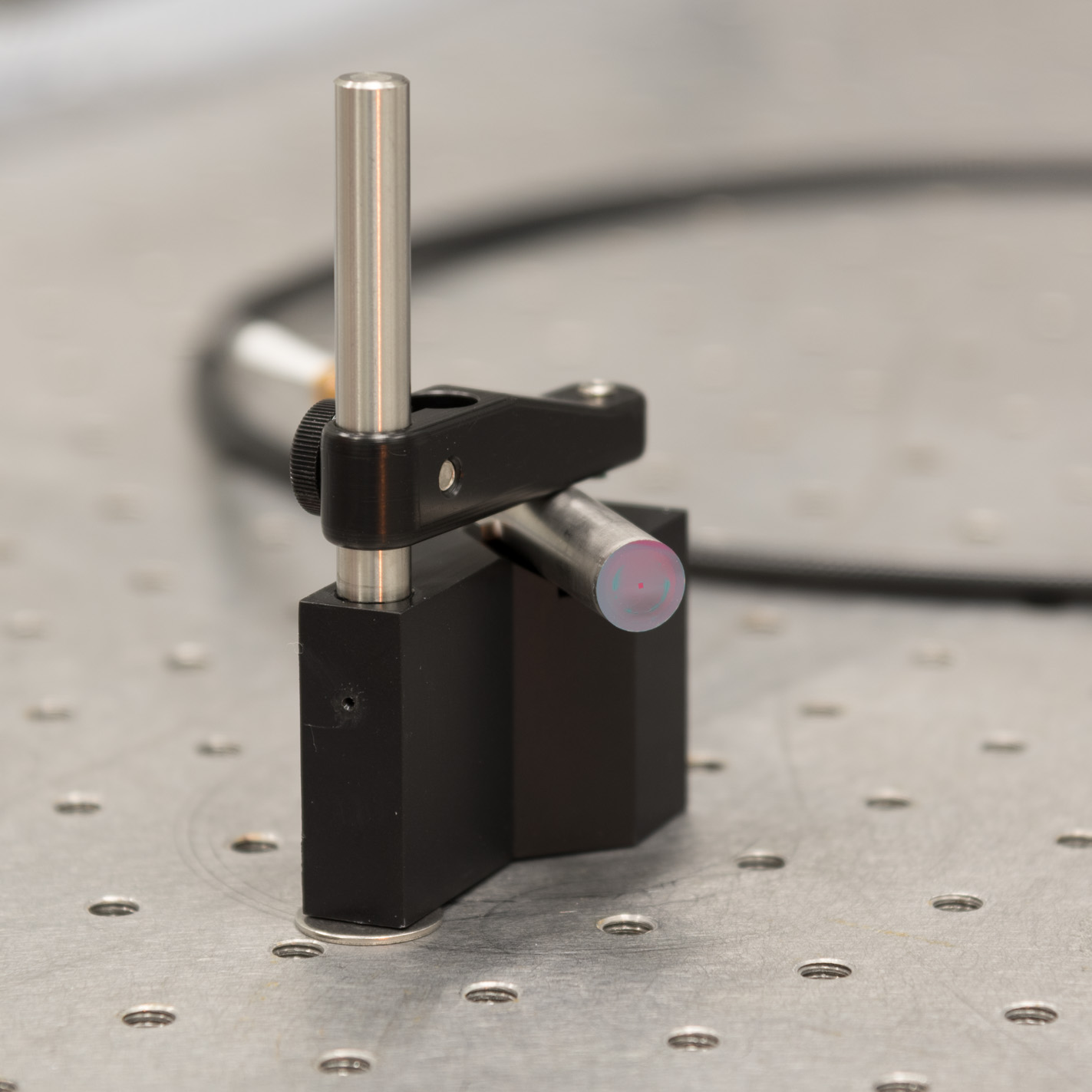}
  \includegraphics[width=0.33\textwidth]{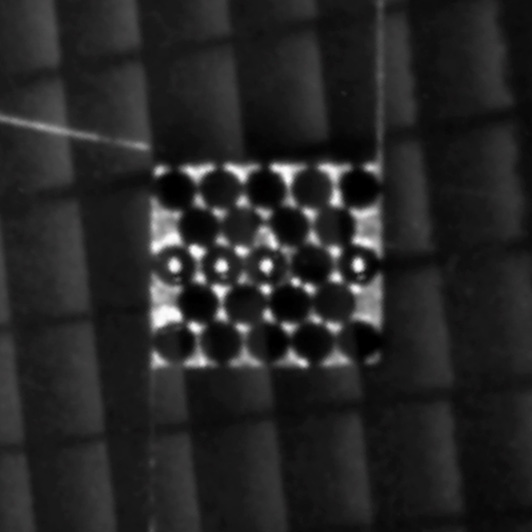}
  \includegraphics[width=0.33\textwidth]{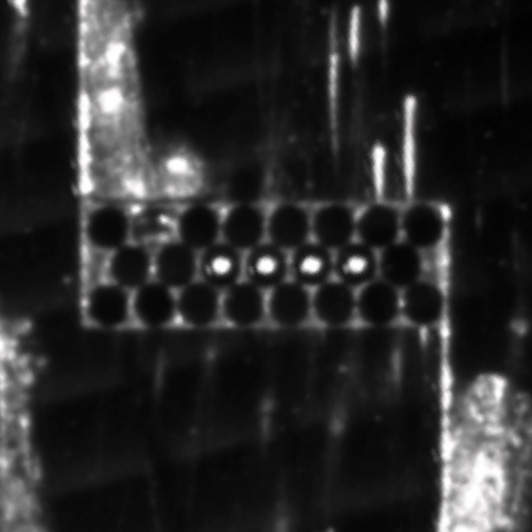}
  \caption{Picture of input ferrule of the FB on the left. The blue and purple colors at the surface of the ferrule are due to the AR coating that was applied to minimize Fresnel losses. The center and right pictures show the input and output patterns of fibers, respectively, which were obtained by imaging the ferrules under a binocular magnifier. In both cases, the science and reference fibers are lit up by illuminating the other ferrule using a halogen lamp. Some reflections and scratches are visible on the surface of both ferrules. The fibers that are not lit up correspond either to calibration fibers that were not illuminated when taking the pictures or to dummy fibers used to pack the geometry of the pattern (see Fig.~\ref{fig:fb_drawing}). The fiber that appears damaged at the bottom right of the output ferrule is a dummy fiber that has no functional purpose.}
  \label{fig:fb_pictures}
\end{figure*}

\begin{figure*}
  \centering 
  \includegraphics[height=8cm]{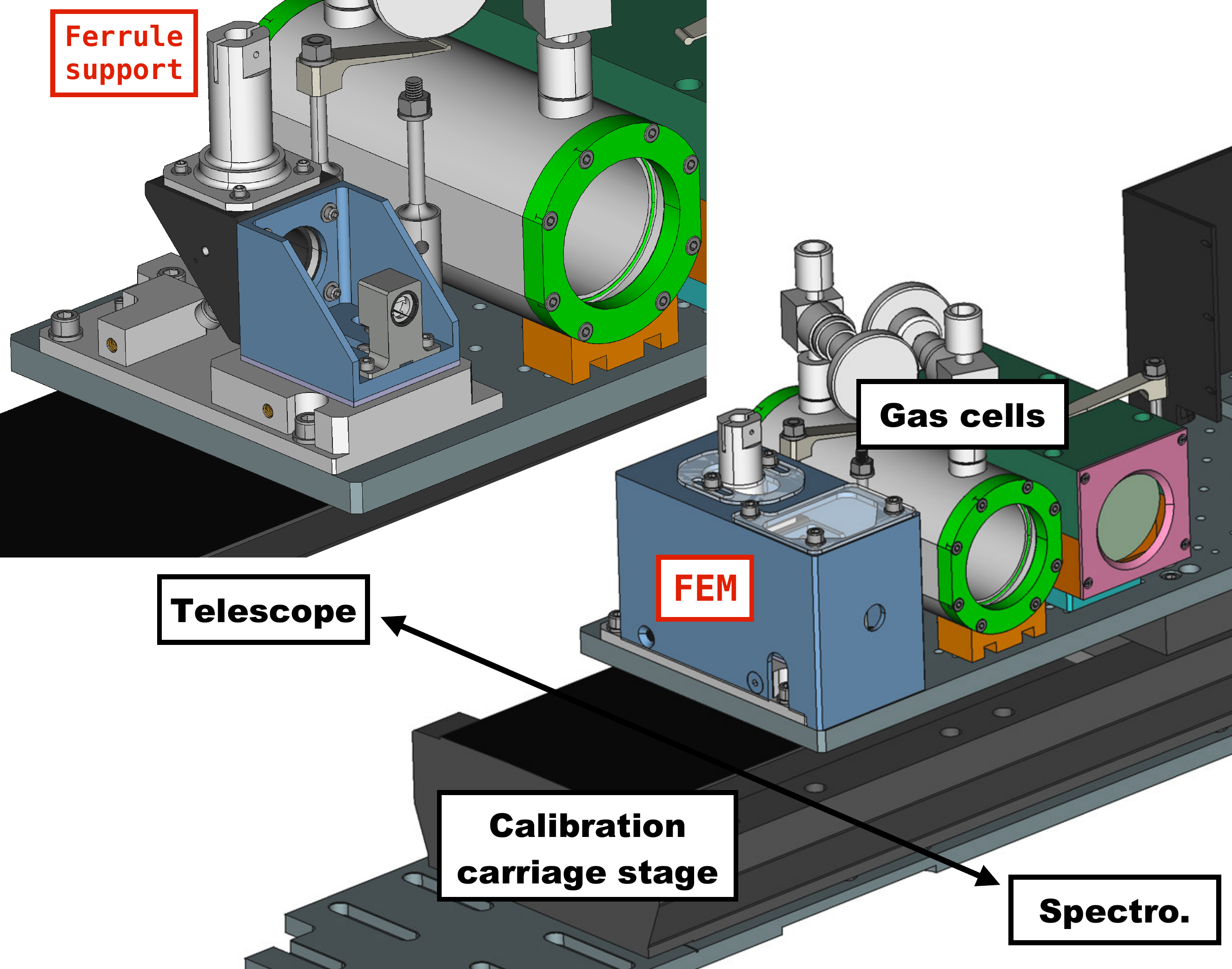}
  \includegraphics[height=8cm]{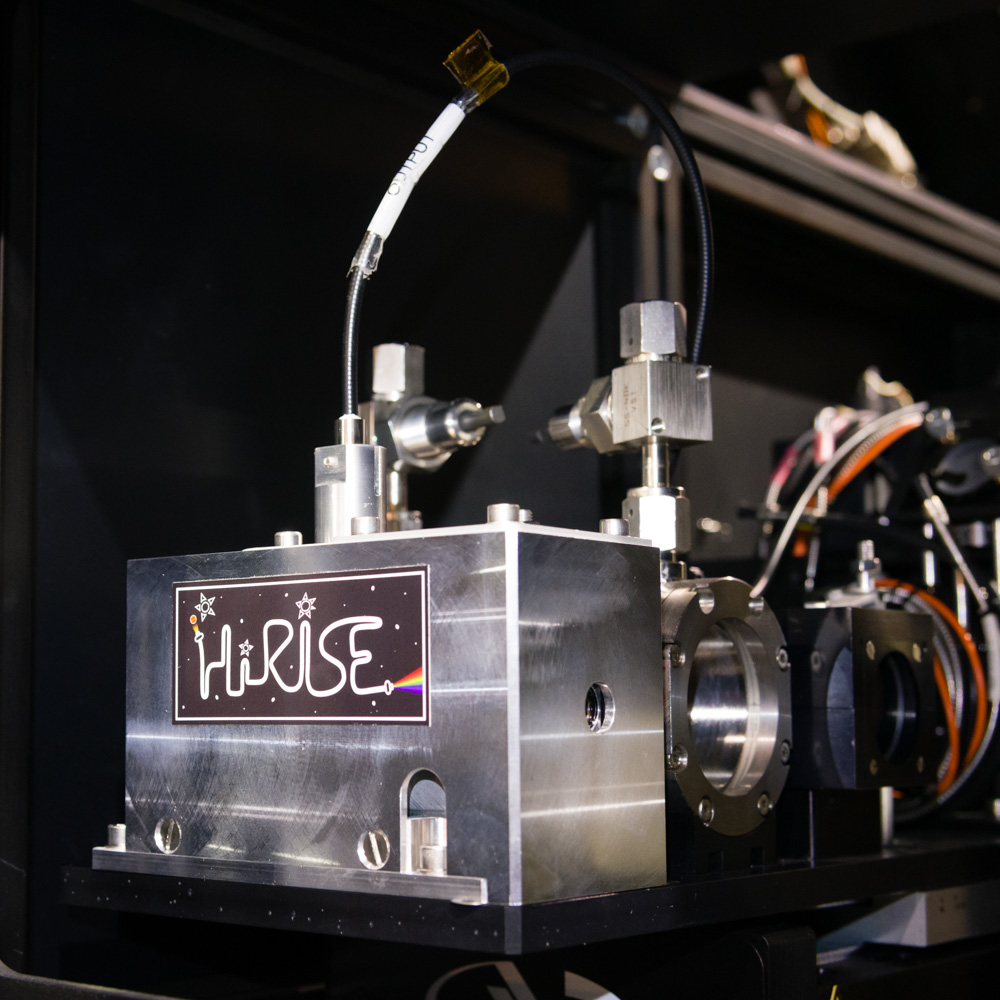}
  \caption{Implementation of the FEM on the calibration carriage stage at the entrance of \crires. In the left drawing, existing \crires components are labelled in black, while the FEM is labelled in red. The inset in the top left shows the inside of the FEM. The output ferrule of the FB is inserted vertically from the top in the ferrule support. The picture on the right shows the FEM after its installation in \crires. The FB (black-coated with a white sticker) is visible over the FEM.}
  \label{fig:fem}
\end{figure*}

The FB of HiRISE is a single piece of hardware that routes science and calibration fibers from SPHERE to \crires. It was manufactured by FiberTech Optica (Canada). It has a total length of 83.9\,m, divided into three sections: a short section in SPHERE (2.5\,m), a long section around the telescope (78.2\,m), and another short section in \crires (3.2\,m). A conceptual drawing of the FB with its input and output ferrules is presented in Fig.~\ref{fig:fb_drawing}.

The choice of science fiber was based on its operating wavelength, its numerical aperture (NA), and its transmission in the  $H$ band. Originally, the optical design of HiRISE was considering off-the-shelf ZBLAN fibers proposed from manufacturer \emph{\emph{Le Verre Fluoré}}, which had $\mathrm{NA} = 0.16$. When it was finally decided to drop the ZBLAN fibers for more standard telecom fibers, the closest match on the market in terms of NA was found to be the Nufern 1310M-HP fiber. The specifications of that fiber are summarized in Table~\ref{tab:science_fiber_specification}. The attenuation of this fiber is $\leq 0.50$\,dB/km at 1.550\,\mic, which translates into an expected transmission $\geq 0.989$ for a 100\,m fiber. Beyond 1.6\,\mic, the attenuation is expected to rise, but laboratory measurements performed during the design phase of HiRISE with a 100\,m sample showed that the transmission remains higher than 0.97 at 1.8\,\mic, which is the upper limit of our spectral range for science.

For science, the bundle has four fibers: one that samples the planetary signal at the center of the FoV (science fiber), and three that sample the PSF or speckle field of the star in the field (reference fibers). These four fibers go all the way from the input ferrule in SPHERE to the output ferrule in \crires. Pictures of the ferrule and input/output fiber arrangements are shown in Fig.~\ref{fig:fb_pictures}. The ferrules are made of Invar to avoid stresses due to temperature variations. In both ferrules, the mechanical center is located within a few micrometers of the center of the science fiber. The fibers are in a packed geometry where they touch, cladding-to-cladding, so the cores are located 125\,\mic apart. To maximize the transmission at the input and output, the ferrules are AR-coated with coatings optimized up to 1.850\,\mic on both sides. The lower limit for the AR-coating is different on the input and output for the reasons detailed below. Taking into account the coating and the fiber, the FB transmits $\sim$95\% of the light that couples into the science fiber (see Sect.~\ref{sec:transmission} for a detailed transmission budget).

\begin{table}[b]
  \caption[]{Specifications of the Nufern 1310M-HP fiber.}
  \label{tab:science_fiber_specification}
  \centering
  \begin{tabular}{rl}
    \hline\hline
    Parameter & Value \\
    \hline
    \multicolumn{2}{c}{Optical specifications} \\
    \hline
    Operating wavelength    & 1310--1620\,nm \\
    Core numerical aperture & 0.16 \\
    Mode field diameter     & $6.7 \pm 0.5$\,\mic @ 1310\,nm \\
                            & $7.6 \pm 0.6$\,\mic @ 1550\,nm \\
    Cutoff wavelength       & $1250 \pm 50$\,nm \\
    Core attenuation        & $\leq 0.75$\,dB/km @ 1310\,nm \\
                            & $\leq 0.50$\,dB/km @ 1550\,nm \\
    \hline
    \multicolumn{2}{c}{Geometrical \& mechanical specifications} \\
    \hline
    Cladding diameter       & $125 \pm 1.0$\,\mic \\
    Core diameter           & 6.0\,\mic \\
    Core/cladding offset    & $\leq 0.5$\,\mic \\
    Bend radius             & $\geq 6$\,mm $[$short term$]$ \\
                            & $\geq 13$\,mm $[$long term$]$ \\
    \hline
  \end{tabular} 
\end{table}

For operations and calibrations on the SPHERE side, the input ferrule includes five calibration fibers. Four of them are the feedback fibers described in Sect.~\ref{sec:fim}, which are connected to an LED light source in the electronics cabinet. The fifth one, called the centering fiber, is connected to a power meter in the cabinet and is used for the target acquisition in the FIM (see Sect.~\ref{sec:operations}).

On the \crires side, two side fibers are placed on the same line as the science and reference fibers to mark the starting and ending positions. These fibers have only been used during AIT at the telescope to help find the orientation of the \crires derotator that aligns the science fibers with the spectrograph's slit. For this purpose, they were temporarily connected to an independent halogen source. Two additional fibers are dedicated to AO calibrations for the \crires MACAO system \citep{Paufique2004}. One is an SMF identical to the science fibers, and the other is an MMF with a 50\,\mic core, $\mathrm{NA} = 0.22,$ and an operating wavelength range of 350--2400\,nm. These guide fibers are used to create a reference for the MACAO wavefront sensor (WFS) and maintain the MACAO deformable mirror (DM) flat during HiRISE observations. More explanations on these aspects are provided in Sect.~\ref{sec:operations}. The MMF fiber was included in case the flux coming from the calibration source through the SMF guide fiber is too low for MACAO, but tests at the telescope showed that the light coming from the SMF is sufficient for MACAO operations.

The fibers are protected within a PVC-coated stainless interlock monocoil conduit. The short SPHERE and \crires sections, which require more flexibility, have an outer diameter of 4.2\,mm, while the long central section has an outer diameter of 9.2\,mm. Inside the long central section the fibers are protected in an additional internal PVC furcation tubing, which is reinforced with Kevlar thread, and a glass fiber tensile core to decrease mechanical stress when the outside temperature changes. The central section has two stainless steel breakouts at both ends, which reinforce the junctions with the short SPHERE and \crires sections. On the SPHERE side, the breakout also lets the ancillary calibration fibers go in and out.

The FB was routed around the UT3, starting on the SPHERE side, using existing cable ducts. The route follows cable ducts underneath the azimuth platforms and then follows the telescope structure to reach the inner track of the telescope under the azimuth floor. An electrical cable with approximately the same minimum radius of curvature as the FB was installed in 2021 to measure the exact length required for the bundle. Another possible shorter route would have been to follow the primary mirror cell, but in that case the bundle would have needed to be included in the altitude cable wrap of the telescope and would have been subject to constant stresses as the telescope is operated. The selected path is therefore slightly longer, but the bundle remains entirely static once it is installed.

\subsection{Fiber extraction module}
\label{sec:fem}

The FEM is a much simpler system than the FIM. It is located in the calibration carrier stage of \crires, which is at the entrance of the warm bench of the instrument. The edge of the carrier stage lies within a couple of centimeters of the VLT focal plane. The warm bench contains the instrument derotator, the MACAO deformable mirror, and the MACAO wavefront sensor \citep{Dorn2023,Paufique2004}. It produces a turbulence-corrected image in the plane of the slit located at the entrance of the spectrograph.

The purpose of the FEM is to reimage the fibers in the output ferrule, which have a focal ratio of F/3.5, into the VLT focal plane at F/15 and with a pupil image at the distance of the UT3 pupil. It is composed of a total of four lenses (including a glued doublet). A flat folding mirror is used to make the system more compact and to ensure that the ferrule is inserted from the top to avoid unnecessary bending of the fibers. All lenses are custom-made by \emph{\emph{Optiques Fichou}} and AR-coated from 0.8 to 1.85\,\mic to cover the science wavelengths as well as the red part of the visible for the calibration of MACAO using one of the AO guide fibers in the FB. The FEM transmits more than 90\% of the light exiting the science fiber (see Sect.~\ref{sec:transmission} for a detailed transmission budget).

The mechanical design of the FEM allows it to fit within the space allocated for visitor gas cells (\ref{TR2}), where it does not impact the regular operations of \crires (\ref{TR1}). Moreover, the FEM has been designed to be easily removed in the case of an observing run requiring a visitor gas cell or during technical interventions. A drawing and picture of the implementation is visible in Fig.~\ref{fig:fem}. Contrary to the FIM, the mechanical parts of the FEM have not been anodized due to a lack of time during AIT.

\section{Operations}
\label{sec:operations}

The operations of HiRISE involve three instruments: SPHERE, the HiRISE FIM, and \crires. In this section we describe the necessary calibrations before HiRISE observations (Sect.~\ref{sec:calibrations}), the science target acquisition (Sect.~\ref{sec:acquisition}), and the science observations (Sect.~\ref{sec:observations}). A summary of the different calibrations and steps is provided in Table~\ref{tab:summary_calibs}.

During commissioning, the operations were done with a mix of existing VLT software templates, a couple of templates developed specifically for HiRISE, and python scripts calling VLT software functions to control hardware. This situation may evolve toward only templates in the future to make the instrument operable by the observatory personnel on a regular basis.

\begin{table*}
  \caption[]{Summary of the daytime calibrations, target acquisition, and science observations.}
  \label{tab:summary_calibs}
  \centering
  \begin{tabular}{lllcc}
    \hline\hline         
    Calibration / Step          & Purpose                                         & Sub-system  & Duration & Section \\
                                &                                                 &             & [min]    & \\
    \hline
    \multicolumn{5}{c}{Daytime calibrations} \\
    \hline
    Tip-tilt mirror linearity   & Monitoring                                      & FIM         &    2 & \ref{sec:calib_sphere_fim} \\
    Interpolation function      & Positioning of the science PSF on the detector  & FIM         &   10 & \ref{sec:calib_sphere_fim} \\
    Interaction matrix          & Positioning of the science PSF on the detector  & FIM         &    2 & \ref{sec:calib_sphere_fim} \\
    Injection map               & Position of centering fiber in tip-tilt space   & FIM         &    2 & \ref{sec:calib_sphere_fim} \\
    Centering optimization      & Position of science fiber in tip-tilt space     & Full system &    2 & \ref{sec:calib_macao_crires} \\
    Fibers trace                & Position of the signal on the science detector  & Full system &    2 & \ref{sec:calib_macao_crires} \\
    \hline         
    \multicolumn{5}{c}{Target acquisition} \\
    \hline
    MACAO acquisition           & Closing the loop on MACAO                       & FIM + MACAO &    2 & \ref{sec:acquisition} \\
    SPHERE acquisition          & Telescope pointing and AO optimization          & SPHERE      &   10 & \ref{sec:acquisition} \\
    Switch to internal source   & FIM acquisition setup                           & SPHERE      & $<$1 & \ref{sec:acquisition} \\
    Injection map               & Position of centering fiber in tip-tilt space   & FIM         &    2 & \ref{sec:acquisition} \\
    Centering optimization      & Position of science fiber in tip-tilt space     & Full system &    2 & \ref{sec:acquisition} \\
    Tracking camera acquisition & Reference position of science fiber on detector & FIM         & $<$1 & \ref{sec:acquisition} \\
    Blind offset                & Positioning companion's PSF on science fiber    & FIM         & $<$1 & \ref{sec:acquisition} \\
    Switch back to sky          & End of FIM acquisition                          & SPHERE      & $<$1 & \ref{sec:acquisition} \\
    Modal gains optimization    & Performance optimization                        & SPHERE      &    2 & \ref{sec:acquisition} \\
    \hline
    \multicolumn{5}{c}{Science observations} \\
    \hline
    Science data acquisition    & Acquisition of spectra                          & \crires     & Hours & \ref{sec:observations} \\
    Continuous tracking         & Maintaining the companion's PSF on S fiber      & FIM         &      & \ref{sec:observations} \\
    \hline
  \end{tabular} 
\end{table*}

\subsection{Calibrations}
\label{sec:calibrations}

A series of calibrations are performed before the beginning of the night to prepare for the observations. All the calibrations below are performed using only the SPHERE internal light sources, which allow them to be done during the afternoon preceding observations. We report the status of calibrations after commissioning here, but we note that they may evolve in the future as we learn more about the instrument.

\subsubsection{Calibrations involving only SPHERE and the FIM}
\label{sec:calib_sphere_fim}

The first calibration is aimed at determining the linear coefficients of the tip-tilt angular motion of the FIM mirror in relation to the PSF motion on the tracking camera detector. The working range of tip-tilt is typically $\sim$4\,mrad with respect to the reference position determined during commissioning. Over this range, the PSF moves by 42.1 and 42.3\,pix/mrad in tip and tilt, respectively. Deviations from linearity remain below 0.2\,pix (0.06\,\loD at $\lambda = 1.6\,\mic$) in the working range.

The second calibration is aimed at determining the coefficient of the interpolation function previously presented by \citet{ElMorsy2022}. The goal of this calibration is to provide a tool that allow us to place the PSF at any desired position on the tracking camera detector. This calibration is performed by recording the position of the PSF for a grid of tip and tilt values on the mirror. The recorded positions are then fed to an interpolation function (currently \texttt{LinearNDInterpolator} from the \texttt{scipy} python package), which returns two functions. These functions take a desired position in pixel coordinates on the tracking camera as a parameter and return the commands in tip and tilt, respectively, to be applied to the mirror to bring the science PSF to the desired location. Immediately after the calibration, the PSF can be placed at the desired location with an accuracy level typically better than 0.1\,pix (0.03\,\loD at $\lambda = 1.6\,\mic$), but temporal drifts related to temperature changes in SPHERE have been observed, which decreases the accuracy. These drifts are on the order of 0.25\,(\loD)/$\degree$C at $\lambda = 1.6\,\mic$ (see Sect.~\ref{sec:stability}), which is significant and requires using a continuous compensation during the science observations (see Sect.~\ref{sec:observations}). A more thorough analysis of the accuracy and stability will be presented in El Morsy et al. (in prep.). The loss of absolute accuracy is not critical for operations, as we explain in Sect.~\ref{sec:acquisition}, which covers the target acquisition.

The third calibration is to build interaction and command matrices for the tip-tilt mirror. The interaction matrix is calibrated by moving it in tip and tilt by a fixed amount of 0.5\,mrad, starting from a position close to that of the science fiber, and recording the motion of the PSF on the tracking camera detector. Then, the command matrix is computed and allows us to move the PSF by small amounts for corrections during target acquisition and observations. The command matrix is used as a complement to the interpolation function described in the previous paragraph. The accuracy of the motions based on the command matrix is typically better than 0.1\,pix (0.03\,\loD at $\lambda = 1.6\,\mic$).

\begin{figure}
  \centering 
  \includegraphics[width=0.48\textwidth]{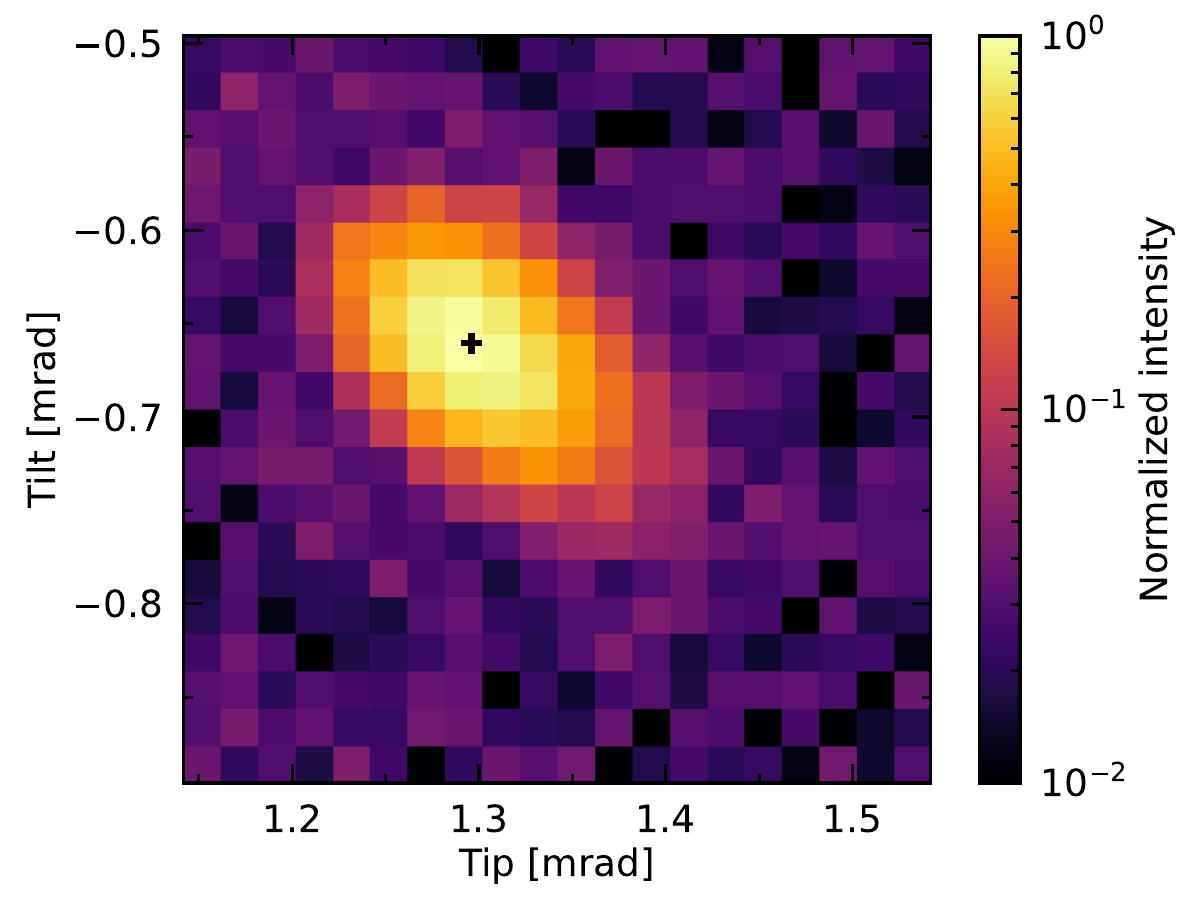}
  \caption{Example of a 20$\times$20 pixel injection map done on the centering fiber with power meter. The sampling of 0.02\,mrad/pixel (0.5\,\loD at $\lambda = 1.6\,\mic$) is relatively coarse but sufficient to obtain an accurate estimate of the center position with a Gaussian fit (black cross). The spot is extended along one of the diagonals because of the projection effect of the pupil on the tip-tilt mirror, which is tilted by 45\degre.}
  \label{fig:injection_map_cen}
\end{figure}

Next, we record an injection map on the centering fiber. Injection maps are obtained by scanning the science PSF in front of a fiber using the tip-tilt mirror and recording the output flux at the other end of the fiber by means of a camera or power meter \citep[e.g.,][]{ElMorsy2022}. The goal is to determine the position of the fiber in tip-tilt space, that is the tip-tilt command that allows the centering of the stellar PSF on the fiber. In HiRISE, we use a centering fiber connected to a power meter inside the electronics cabinet. Power meters offer the advantage of a very fast readout ($< 1$\,ms) and relatively good sensitivity in the $H$ band with InGaAs photodiodes. While this sensitivity is not sufficient to perform on-sky injection maps, except for extremely bright stars, it is sensitive enough to work on the SPHERE internal source. For the daily calibration on the centering fiber, we sample a grid of 0.4$\times$0.4\,mrad$^2$ 20 times in each dimension (400 positions in total), which produces an injection map in less than 1\,min and provides a good level of accuracy on the position of the fiber. An example of a daily injection map obtained during commissioning is presented in Fig.~\ref{fig:injection_map_cen}.

\subsubsection{Calibrations involving MACAO and \crires}
\label{sec:calib_macao_crires}

\begin{figure}
  \centering 
  \includegraphics[width=0.5\textwidth]{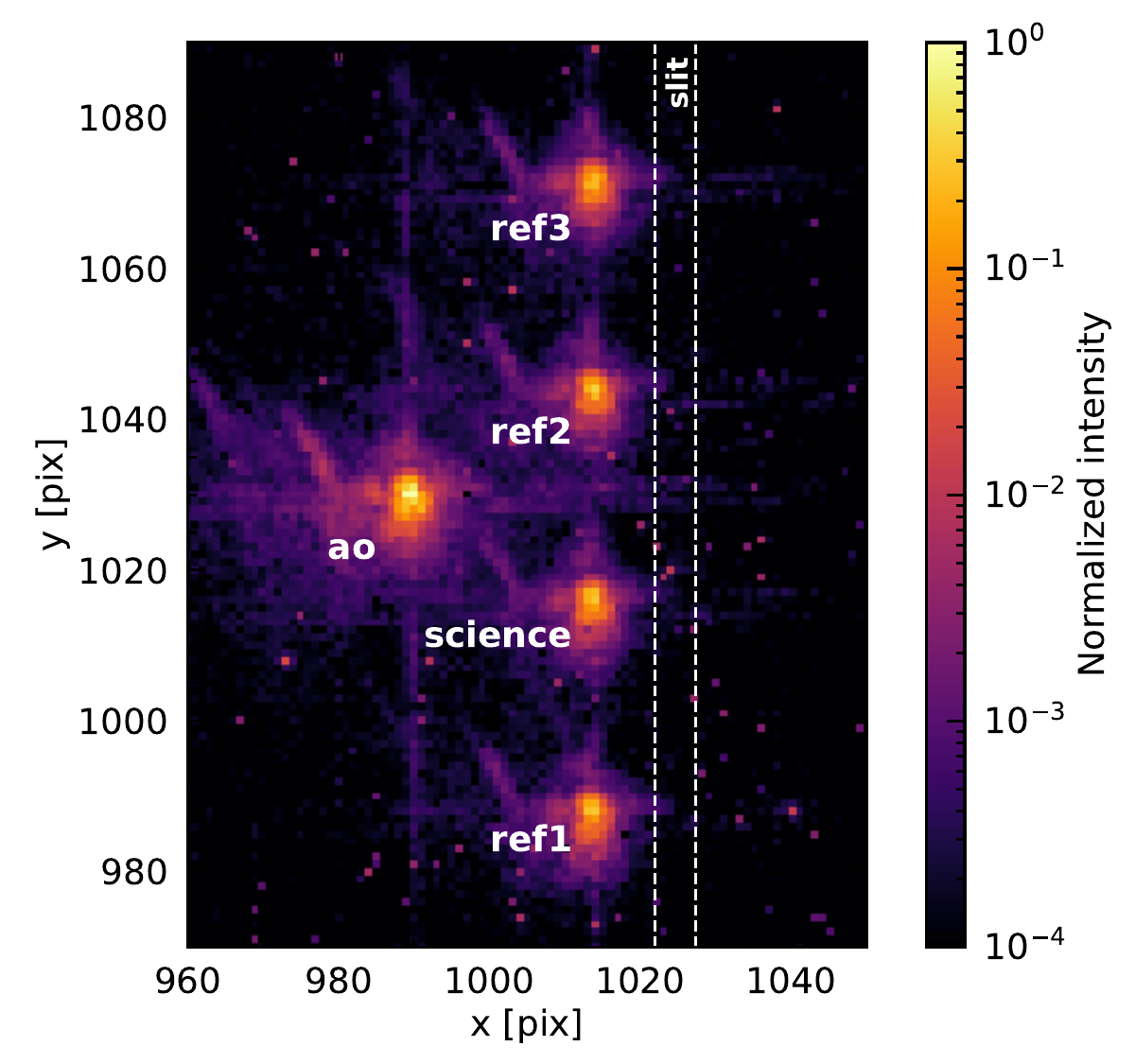}
  \caption{Slit viewer (SV) camera image of fibers reimaged by the FEM. The cross-like diffraction pattern, particularly visible for the AO guide fiber, is induced by a pupil mask installed on the MACAO DM for calibrations. The image is done with a closed loop on MACAO and by offsetting the center of the guide window on the SV camera so that the science and reference fibers are moved out of the slit (dashed vertical lines).}
  \label{fig:sv_fibers}
\end{figure}

\begin{figure*}
  \centering 
  \includegraphics[width=1\textwidth]{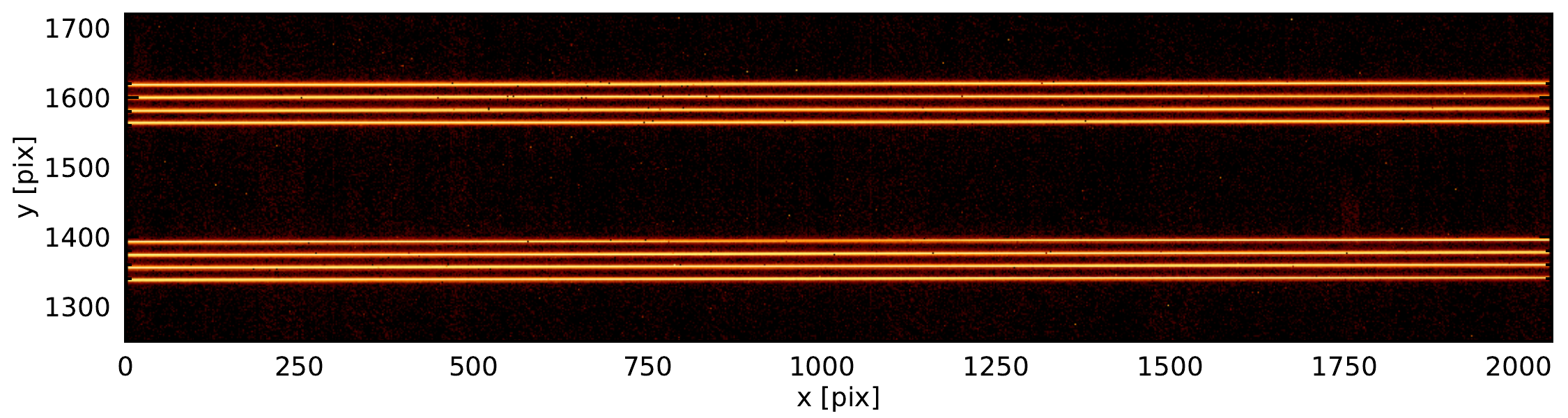}
  \caption{Portion of a trace image on \crires science detector \#1, showing third and fourth spectral orders. The image was obtained by illuminating the science and reference fibers with a flat-field source in SPHERE. This image is used to measure the position the fibers in the spectral data to facilitate signal extraction.}
  \label{fig:crires_trace}
\end{figure*}

With the FEM at the entrance of \crires, the HiRISE light goes through the entire instrument (warm bench and spectrograph), including the MACAO AO system. Because of its intrinsic properties, the MACAO DM cannot remain completely flat for long periods of time and must be used in a closed loop. This is why the FB includes two guide fibers (one SMF, one MMF) fed by a halogen source: one of these fibers is used as a guide ``star'' for the MACAO WFS to work in a closed loop and keep the DM flat and stable during the observations. After some testing during commissioning, the SMF was considered sufficient in terms of flux, and it is now the reference fiber for operations of HiRISE. A picture of the AO guide fiber, fed by the halogen source, and the science and reference fibers, illuminated by a flat-field source in SPHERE, is shown in Fig.~\ref{fig:sv_fibers}. Since everything is completely static during HiRISE observations, the MACAO parameters for this guide fiber have been determined once and do not require regular recalibration. For this reason, HiRISE daily calibrations do not require any MACAO-specific calibrations.

The \crires control software also includes a secondary guiding based on the position of the guide star on the slit viewer (SV) camera. This guiding ensures that any differential motion between the warm bench and the cold spectrograph, which are not physically tied, is compensated during observations. For HiRISE, the center of the guide window has been calibrated so that the science and reference fibers fall exactly at the center of the 0.2\as slit. Again, the secondary guiding parameters are not expected to vary over time, so they were only calibrated once during commissioning.

The SV secondary guiding offers the possibility to update the center of the guide window in real time to offset the science and reference fibers from the slit, and make them become visible on the SV camera. Figure~\ref{fig:sv_fibers} is obtained in such a configuration, with the center of the guide window offset by eight pixels from its reference position. In this configuration, we calibrate the position of each fiber for flux measurements every day (see below) by fitting a 2D\ Gaussian function on the spots. These values are stored in a local database for later use during the calibrations and observations.

From the data of Fig.~\ref{fig:sv_fibers}, we can estimate the fraction of light that goes through the slit: 89\% for the 0.2\as slit (baseline for HiRISE observations) and 96\% for the 0.4\as slit. For the 0.2\as slit, this estimation is in good agreement with the estimation of 92\% from the optical design.

In the offset configuration, it is possible to produce injection maps on the science fiber by measuring the output flux on the SV camera. This is, however, much slower than for the centering fiber because it involves several more steps than simply reading the output value of a power meter: image acquisition on the slit viewer (typically 0.5 to 1\,s), image transfer to the HiRISE workstation ($\sim$1\,s), image reading and analysis ($\sim$1\,s). This is why this solution is not used in the daily calibrations and was only used a few times during commissioning to check its feasibility. Instead, we use the calibrated offset between the centering and science fibers ($-1.483$\,mrad in tip, $+1.551$\,mrad in tilt) to place the PSF on the science fiber after the injection map on the centering fiber. Then, we run a Nelder-Mead optimization using the \texttt{scipy.optimize} package to maximize the flux measured at the output of the science fiber on the SV camera, which we use as a proxy to maximize the injection in the fiber. The optimization is limited to ten iterations, or 20 evaluations of the flux on the SV camera, to minimize the time of the calibration. Although some additional testing would need to be done, we do not find that a higher number of iterations provides a higher output flux, which seems to indicate that the centering in ten iterations is accurate enough. The time required for this calibration is of the order of two minutes: 1 to 1.5 minutes for the optimization algorithm, and two times 15\,s to introduce and remove the offset on the SV guide window.

Finally, the last calibration involves applying the flat-field source in SPHERE to illuminate the science and reference fibers (like in Fig.~\ref{fig:sv_fibers}), but this time with an image of the fibers falling in the slit and being fed into the spectrograph. This allows the production of a ``trace image'' where the location of each fiber for each order and detector is clearly identifiable. This calibration will later be used in the data analysis when extracting the flux of a companion. A portion of a trace image is visible in Fig.~\ref{fig:crires_trace}.

\subsection{Target acquisition}
\label{sec:acquisition}

The science target acquisition is a multistep process. At the beginning of the night, a MACAO specific template is executed on the HiRISE workstation to set MACAO up, acquire a background on the SV camera, and close the MACAO and the secondary guiding loops on the HiRISE AO guide fiber. With \crires being used with the fixed constellation of HiRISE fiber bundle, all sky-related controls are made static: derotator angle; off-axis AO tracking; field stop size and position. In principle, this template can be executed once per night because the MACAO configuration is static and the loop can remain closed on the guide fiber for many hours without any disturbance. However, as we see in Sect.~\ref{sec:performance}, a different operational scheme may be necessary in the future. The execution of this template, which as been developed specifically for HiRISE, takes less than a minute.

Then, the SPHERE \texttt{IRDIFS-EXT} acquisition template is executed. Briefly speaking, this template sends a preset to the telescope, set SPHERE up for observation, and optimizes the SPHERE ExAO system (SAXO). Since we do not use a coronagraph with HiRISE to maximize the number of planetary photons that reach the spectrograph's detector, the acquisition template does not perform any of the additional steps that are typically required when using a coronagraph such as optimizing focus and centering. It typically takes between five and ten minutes depending on telescope preset time and observing conditions.

The final, most critical step is the centering of the substellar companion's PSF on the science fiber at the level of the FIM. This step is difficult because part of the error budget is tied to the uncertainties of the on-sky astrometry of the companion coming from previous observations, over which we have no control. It is therefore important to be as accurate as possible on the centering with the FIM to maximize the injection efficiency into the science fiber.

For better performance, the centering is performed on the bright internal source of SPHERE, which provides a highly stable and repeatable PSF. After the target acquisition in SPHERE, we immediately switch back to the internal source and perform a fast injection map on the centering fiber to find its approximate location. Then, we offset the guide window of the secondary guiding on the SV camera of \crires to move the science and reference fibers out of the slit, we apply the calibrated offset on the FIM tip-tilt mirror to place the source's PSF on the science fiber, and we perform an optimization to accurately center the PSF. Finally, we remove the secondary guiding offset to move back the fiber to the slit. After these steps, we know that the internal source's PSF is accurately centered on the science fiber.

At this stage, a tracking camera image is acquired to define the corresponding location of the science fiber on the detector. From this and the known astrometry of the companion, we compute the detector location at which the stellar PSF must be placed so that the companion's PSF is centered on the science fiber. The PSF is then moved to that location using the interpolation function (see Sect.~\ref{sec:calibrations}). Because of the temporal drifts, the PSF usually falls within one pixel of the requested location, so we implement a correction loop using the command matrix to refine the position. The loop is stopped when the measured position of the PSF is within 0.05\,pixel (0.015\,\loD at $\lambda = 1.6\,\mic$) of the requested position, or when a maximum of ten iterations have been performed.

Finally, SPHERE is switched back to skylight to reacquire the science target. A pause of 90 seconds is added to wait for the ExAO modal gains to be updated to values adequate for the current on-sky conditions. Indeed, when switching to the internal source the modal gains automatically update to a very small value, of the order of 0.01, because there is no turbulence to correct.

After switching back to sky, we rely on a major feature of the SPHERE ExAO system to ensure that the companion's PSF will actually be centered on the science fiber: the differential tip-tilt sensor \citep{Beuzit2019}. This sensor is located just before the coronagraphic focal-plane masks in the NIR arm of SPHERE. It picks up a very small fraction of light near the H-band to produce a PSF on a 32$\times$32\,pix detector. A dedicated control loop, working at 1\,Hz, ensures that the PSF remains stable at a position defined by calibrated reference slopes. The classical role of the DTTS loop is to keep the PSF perfectly stable on the focal-plane mask.

For HiRISE, we divert this feature to our advantage; if the internal source of SPHERE is centered on a given DTTS pixel, then the DTTS loop ensures that the on-sky stellar PSF is centered identically on sky. This means that after the full FIM centering procedure on the internal source, the stellar PSF will still fall at the same exact location of the FIM tracking camera, and the PSF of the substellar companion will still be centered on the science fiber.

\subsection{Observations}
\label{sec:observations}

The science observations are more straightforward. They are executed using a HiRISE specific template that runs directly on the HiRISE workstation and sends commands to the \crires workstation for data acquisition.

At the beginning of the template, the internal \crires metrology is run for the user-requested spectral setting. This ensures a more accurate wavelength calibration over long observations. The use of the metrology is highly recommended by the user manual for any \crires observations (see \citealt{Dorn2023} for details on the metrology). The metrology is automatically run in case of a change of spectral setting with respect to the previous observation, or if the previous execution was done more than one hour before.

HiRISE then sets \crires in the relevant configuration. SPHERE remains in control of all telescope motions, while \crires keeps --internally-- the relevant secondary loops active. In particular, the SV keeps sending offsets to MACAO, while differential refraction correction is not active. The offsets are performed with combined offsets of MACAO field selector and of its tip-tilt mount, as well as of the SV camera guide window. Although not mandatory, small offsets between exposures can be used to avoid having the spectra always falling on the same bad pixels of the science detectors. 

Finally, science exposures are acquired with a detector integration time (DIT) specified by the user. The user has the possibility to choose the number of DIT per exposure (NDIT) and the number of exposures per offset position (NEXP).

\begin{figure}
  \centering
  \includegraphics[width=0.48\textwidth]{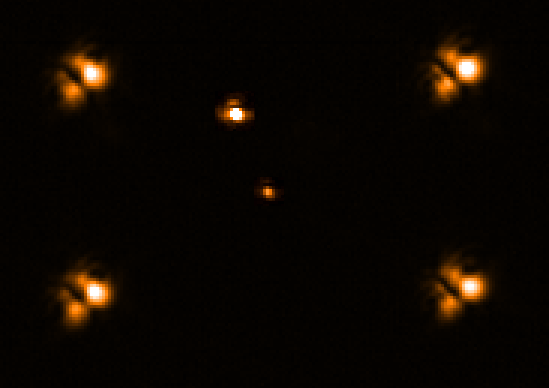}
  \caption{Screenshot of the tracking camera image obtained after centering a stellar companion on the science fiber (roughly at the center of the image). The primary star is located in upper left of the companion. The four broad spots forming a square are the images of the feedback fibers. Their specific size and shape are due to the optics in the feedback channel of the FIM, and in particular to the use of a corner cube for the retro-reflection. The center of each feedback fiber is obtained with a 2D Gaussian fit on the brightest peak of their image.}
  \label{fig:fiber_tracking}
\end{figure}

During long science observations, a tracking loop is run every minute to maintain the companion's PSF aligned as accurately as possible with the science fiber (Fig.~\ref{fig:fiber_tracking}). This loop uses the image of the feedback fibers (Sect.~\ref{sec:fb}) on the tracking camera, through the dedicated FIM feedback channel (Sect.~\ref{sec:fim}). AIT and commissioning measurements have demonstrated that the position of the feedback spots is well correlated with the position of the image of the centering fiber on the tracking camera detector (see analysis and caveats in Sect.~\ref{sec:stability} below). 

The loop therefore tries to ensure that the image of the stellar PSF remains stable with respect to the crossing point $\mathcal{I}$ of the lines going through opposite feedback points. When a measurement is made in the loop, the distance between $\mathcal{I}$ and the stellar PSF is measured and compared to the distance measured immediately after the target acquisition. If the distance is above a predefined threshold (0.05\,pix during commissioning), the FIM tip-tilt mirror is moved using the command matrix to bring the PSF back to the calibrated distance, in a maximum of three iterations. The parameters of this tracking loop will be adjusted in the future as more experience is gained with HiRISE.

\section{Performance}
\label{sec:performance}

\subsection{Astrometry}
\label{sec:astrometry}

\begin{figure}
  \centering
  \includegraphics[width=0.48\textwidth]{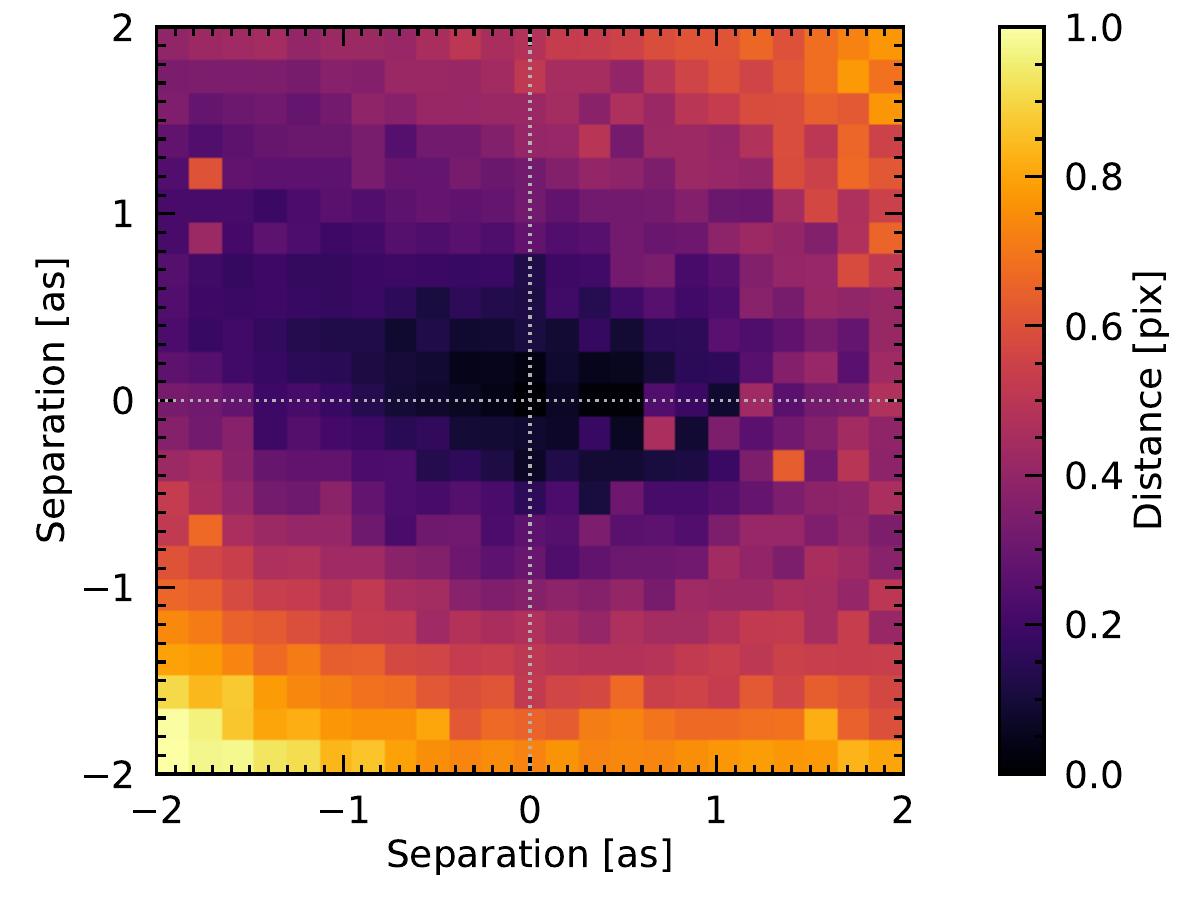}
  \caption{Distortion map on the FIM tracking camera measured using the distortion grid of the SPHERE calibration unit. The analysis covers only a 2\as$\times$2\as FoV around the field center because this is the field for which HiRISE is optimized.}
  \label{fig:distortion_map}
\end{figure}

\begin{table*}
  \caption[]{Astrometric calibration of HiRISE.}
  \label{tab:astrometric_calibration}
  \centering
  \begin{tabular}{p{2.2cm}|c|cc|cc|cc}
    \hline\hline
                &            & \multicolumn{2}{c|}{IRDIS}            & \multicolumn{2}{c|}{HiRISE detector} & \multicolumn{2}{c}{HiRISE sky} \\
    Target      & Epoch      & Sep. sky          & PA sky            & Sep.             & PA                & Pixel scale        & North correction \\
                & [mjd]      & [mas]             & [deg]             & [pixel]          & [deg]             & [mas/pixel]        & [deg] \\
    \hline
    HIP\,103311 & 60137.2326 & $774.11 \pm 0.25$ & $317.08 \pm 0.04$ & $60.28 \pm 0.03$ &~~$47.35 \pm 0.16$ & $12.842 \pm 0.008$ & $90.27 \pm 0.31$ \\
    HIP\,109344 & 60137.2498 & $701.04 \pm 0.23$ &~~$22.06 \pm 0.04$ & $54.69 \pm 0.03$ & $112.44 \pm 0.16$ & $12.818 \pm 0.008$ & $90.38 \pm 0.21$ \\
    HIP\,65288  & 60137.9681 & $761.39 \pm 0.25$ & $245.17 \pm 0.04$ & $59.67 \pm 0.02$ & $335.53 \pm 0.15$ & $12.760 \pm 0.006$ & $90.36 \pm 0.04$ \\  
    HIP\,77939  & 60138.0090 & $362.80 \pm 0.12$ & $111.99 \pm 0.04$ & $28.39 \pm 0.04$ & $202.40 \pm 0.16$ & $12.779 \pm 0.019$ & $90.41 \pm 0.08$ \\
    HIP\,82460  & 60138.0694 & $805.10 \pm 0.26$ & $170.27 \pm 0.04$ & $62.81 \pm 0.04$ & $260.64 \pm 0.16$ & $12.818 \pm 0.009$ & $90.37 \pm 0.06$ \\
    HIP\,95925  & 60138.2418 & $786.74 \pm 0.26$ &~~$29.48 \pm 0.04$ & $61.44 \pm 0.03$ & $120.06 \pm 0.16$ & $12.805 \pm 0.008$ & $90.58 \pm 0.17$ \\
    HIP\,114382 & 60138.3045 & $823.06 \pm 0.27$ & $149.19 \pm 0.04$ & $64.36 \pm 0.07$ & $239.49 \pm 0.16$ & $12.788 \pm 0.015$ & $90.30 \pm 0.07$ \\
    HIP\,116880 & 60138.4111 & $690.17 \pm 0.23$ & $195.24 \pm 0.04$ & $53.82 \pm 0.04$ & $285.76 \pm 0.16$ & $12.824 \pm 0.010$ & $90.52 \pm 0.05$ \\
    \hline
    \multicolumn{6}{l}{Weighted mean}                                                                       & $12.805 \pm 0.027$ & $90.40 \pm 0.08$ \\
    \hline
  \end{tabular} 
\end{table*}

The astrometric calibration is a crucial aspect for HiRISE because, in most cases, we rely on known relative astrometry of substellar companions with respect to their host star to place their PSF on the science fiber. For HiRISE, the astrometric calibration relies on close binaries ($< 1.5\as$) observed in parallel with the FIM tracking camera and with SPHERE/IRDIS. The IRDIS astrometric strategy and accuracy have been determined in the SPHERE GTO and are well documented \citep{Maire2021}. The cross-calibration with IRDIS is therefore the most straightforward option for HiRISE. Moreover, most astrometric fields are too faint or too extended to be observed with the FIM tracking camera.

During commissioning we observed a total of eight binaries that were recalibrated in separation and position angle using IRDIS, and we used them to determine the pixel scale and north orientation of the HiRISE tracking camera. IRDIS images are obtained in the $K$ band with K1 dual-band imaging filter of the K12 pair \citep{Vigan2010}. They are corrected for the $1.0075 \pm 0.0004$ anamorphic distortion, and the position of the two components of each binary is fit with a 2D Gaussian. Based on these positions, we computed the on-sky separation and position angle using the known $12.258 \pm 0.004$\,mas/pix pixel scale in K1 and $-1.76\degre \pm 0.04\degre$ true north correction for IRDIS \citep{Maire2021}. FIM images are obtained on the tracking camera using a series of 17 offsets with the tip-tilt mirror to pave the field. Similarly to IRDIS images, position of the two components of the binaries and determined using a 2D Gaussian, and the detector separation and position angle are computed. Finally, the pixel scale and true north correction are derived for each binary. All the commissioning values are summarized in Table~\ref{tab:astrometric_calibration}.

As IRDIS, the FIM is affected by an anamorphic distortion due to the SPHERE toric mirrors \citep{Hugot2012} located upstream in the optical path. We calibrated the amplitude of the anamorphosis with the distortion grid of the SPHERE calibration unit \citep{Wildi2010} by comparing the distances between all pairs of points along the horizontal and vertical directions. With this method, we estimate a scale difference of a factor $0.42\% \pm 0.10\%$ between the horizontal and vertical directions. To correct for this effect, the FIM images must be extended by a factor 1.0042 along the vertical direction. The distortion map measured on the FIM tracking camera is illustrated in Fig.~\ref{fig:distortion_map}. It is dominated by the anamorphic effect, but higher order distortion effects are also noticeable. A more refined astrometric correction will be built over time and implemented for future observations.

We used a weighted mean to combine the pixel scale and north correction values from Table~\ref{tab:astrometric_calibration}. The weights used in the combination are inversely proportional to size of the error bar for each data point, which gives less weight to the points with larger error bars. We finally determine values of $12.805 \pm 0.027$\,mas/pixel for the pixel scale and $90.40\degre \pm 0.08\degre$ for the true north correction. These values will of course be consolidated and refined in the future with new observations of binaries as part of the long-term calibration plan of HiRISE.

\subsection{Stability}
\label{sec:stability}

\begin{figure}
  \centering 
  \includegraphics[width=0.48\textwidth]{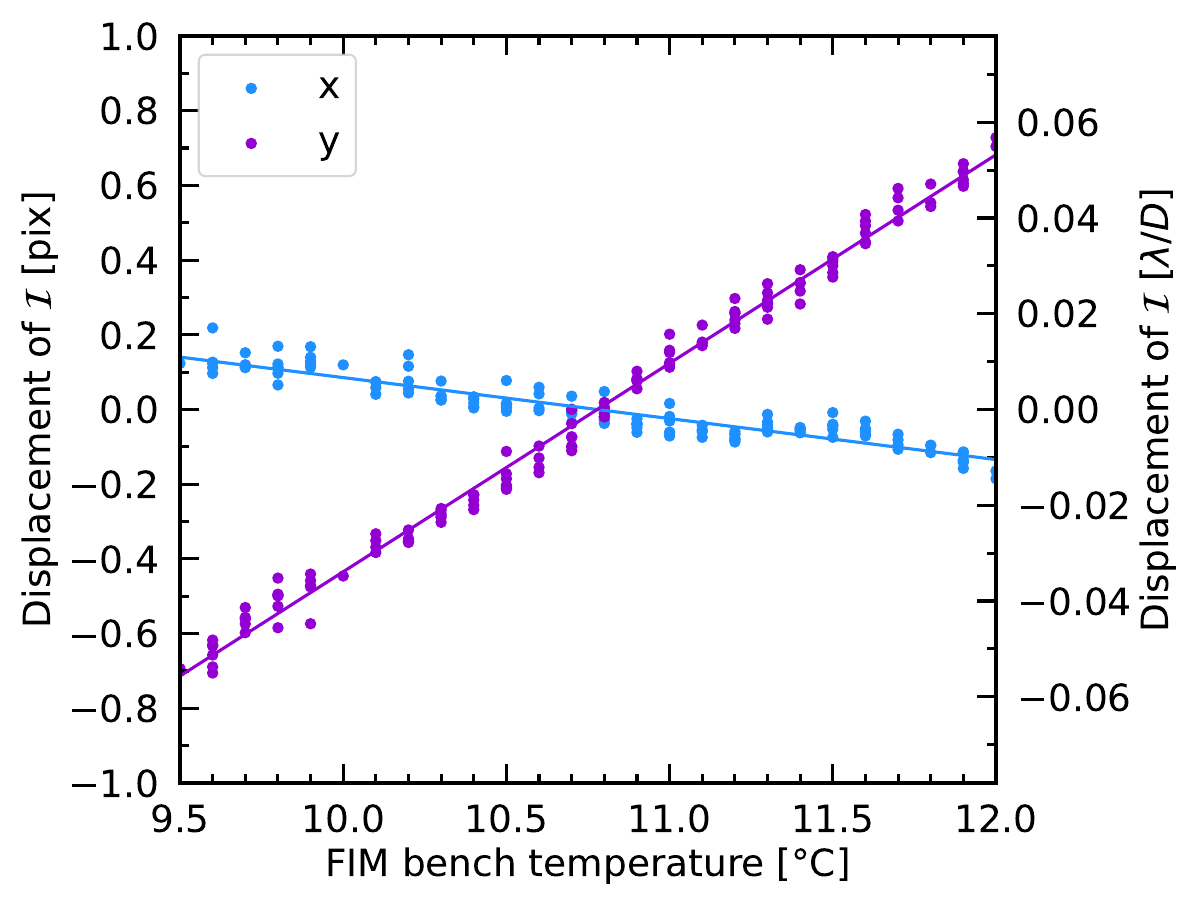}
  \caption{Stability of feedback fibers' crossing point $\mathcal{I}$ with the FIM bench temperature. $\mathcal{I}$ is computed as the crossing point, on the tracking camera detector, between the two lines going through the image of opposite feedback fibers.}
  \label{fig:feedback_stability}
\end{figure}

\begin{figure}
  \centering 
  \includegraphics[width=0.48\textwidth]{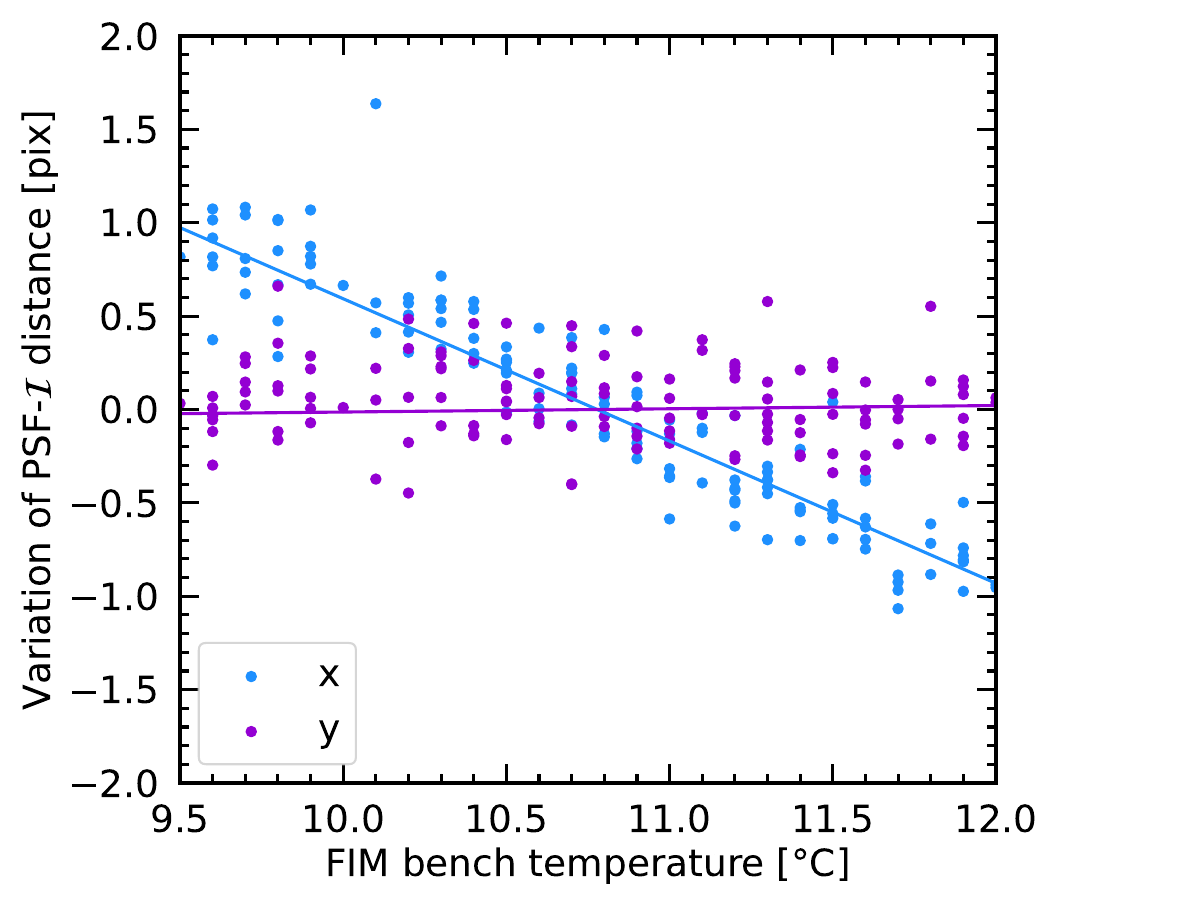}
  \caption{Stability of the distance between $\mathcal{I}$ and the PSF with temperature. In this test the PSF location is used as a proxy for the position of the centering fiber, because before each acquisition the PSF is centered on the fiber using an optimization.}
  \label{fig:feedback_cen_stability}
\end{figure}

The FIM is particularly sensitive to changes in temperature inside the SPHERE enclosure because the bench is $\sim$80\,cm high and made of aluminum. It is expected that the variations of temperature will make the bench shrink or expand, especially when taking into account that the SPHERE enclosure is not stabilized in terms of temperature. The temperature inside the enclosure follows the outside temperature with a delay of 30 to 60 minutes. The temperature gradients are, in theory, on the order of 0.25\degreC/h at most within the enclosure, but in practice steeper gradients up to 0.75\degreC/h were observed during some of our tests.

During commissioning we studied the stability of the fibers with respect to the detector of the tracking camera with a dedicated test. In this test, every minute for several hours, the PSF of the SPHERE internal source is centered on the centering fiber using the optimization described in Sect.~\ref{sec:calibrations}; then, an image of the PSF is acquired with the tracking camera to record its position (using a 2D Gaussian), and finally another image with the tracking fibers illuminated is also acquired to record their position (using a 2D Gaussian). At each iteration of the data acquisition, the bench temperature and timestamps are recorded. The centering fiber is used instead of the science fiber because the centering optimization is much faster on that fiber. The fibers are all packed together in the bundle, so there is no reason to assume the possibility of a differential motion between the science and centering fibers.

The result of this test is presented in Fig.~\ref{fig:feedback_stability}. Instead of showing the displacement of all four feedback fibers on the detector with temperature, we used the crossing point $\mathcal{I}$ as a proxy. $\mathcal{I}$ is simply computed as the crossing point between lines going through the image of opposite feedback fibers. There is an almost perfect correlation with temperature in both the x and y directions, with Pearson correlation coefficients of $-0.933$ and $0.996,$ respectively. The combined displacement of $\mathcal{I}$ on the detector is of 0.79\,pix/$\degree$C, or 0.25\,(\loD)/$\degree$C at $\lambda = 1.6\,\mic$. This displacement is compatible with the length of the optical beam and the temperature expansion coefficients of aluminum.

We also analyze the variation of the distance between $\mathcal{I}$ and the PSF with temperature in Fig.~\ref{fig:feedback_cen_stability}. Here, the PSF is used as a proxy for the centering fiber since at each iteration the PSF is centered using an optimization. Ideally, we hope that the distance remains perfectly constant with temperature. The data are much noisier, with a standard deviation of the residuals of the order of 0.5\,pixel (0.16\,(\loD)/$\degree$C at $\lambda = 1.6\,\mic$), which is probably due to the intrinsic accuracy on the centering fiber. The distance appears stable in y, with a slope of $+0.02$\,pix/$\degree$C ($+0.01$\,(\loD)/$\degree$C at $\lambda = 1.6\,\mic$), but it is not the case in x, with a slope of $-0.76$\,pix/$\degree$C ($-0.24$\,(\loD)/$\degree$C at $\lambda = 1.6\,\mic$).

These results indicate that the loop described in Sect.~\ref{sec:observations} is not perfectly accurate to compensate for the decentering of the PSF on the science fiber with temperature. For a more accurate correction, the temperature variations will need to be taken into account in the loop to remain within the centering specification of 0.1\,\loD recommended by \citet{ElMorsy2022}. This will be investigated in the future after further testing with the system.

\subsection{Non-common path aberrations}
\label{sec:ncpa}

\begin{figure}
  \centering 
  \includegraphics[width=0.48\textwidth]{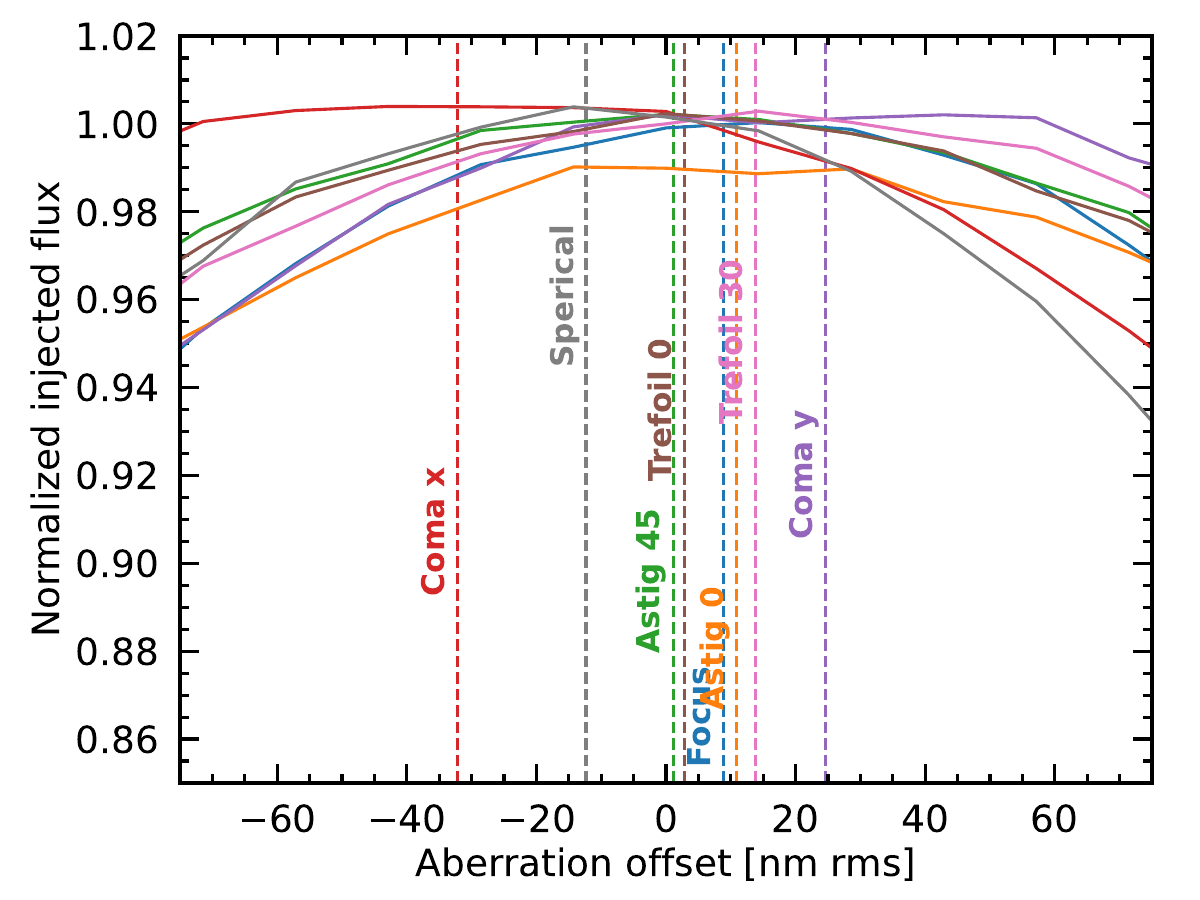}
  \caption{NCPA measured at level of science fiber using aberration offsets applied as Zernike modes to the SPHERE deformable mirror (see text for details). The analysis was limited to focus, astigmatisms, comas, trefoils, and spherical aberration.}
  \label{fig:ncpa}
\end{figure}

\begin{figure*}
  \centering 
  \includegraphics[width=1\textwidth]{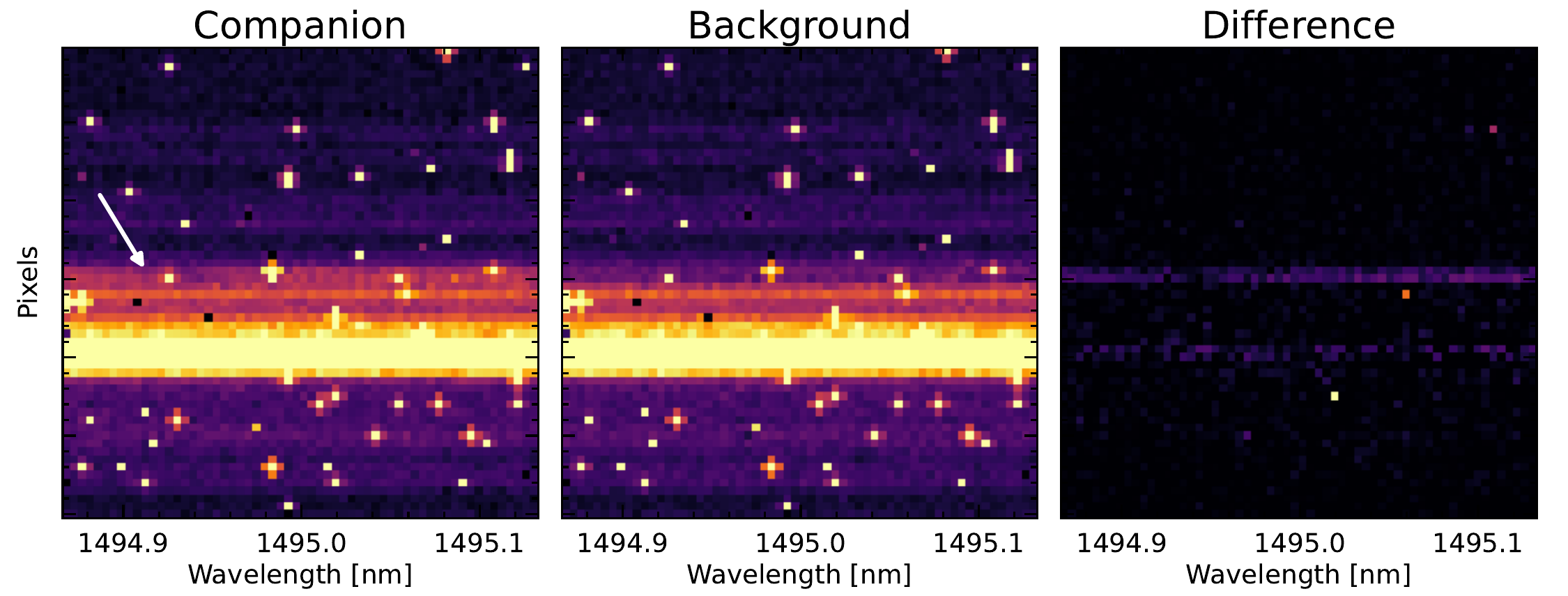}
  \includegraphics[width=1\textwidth]{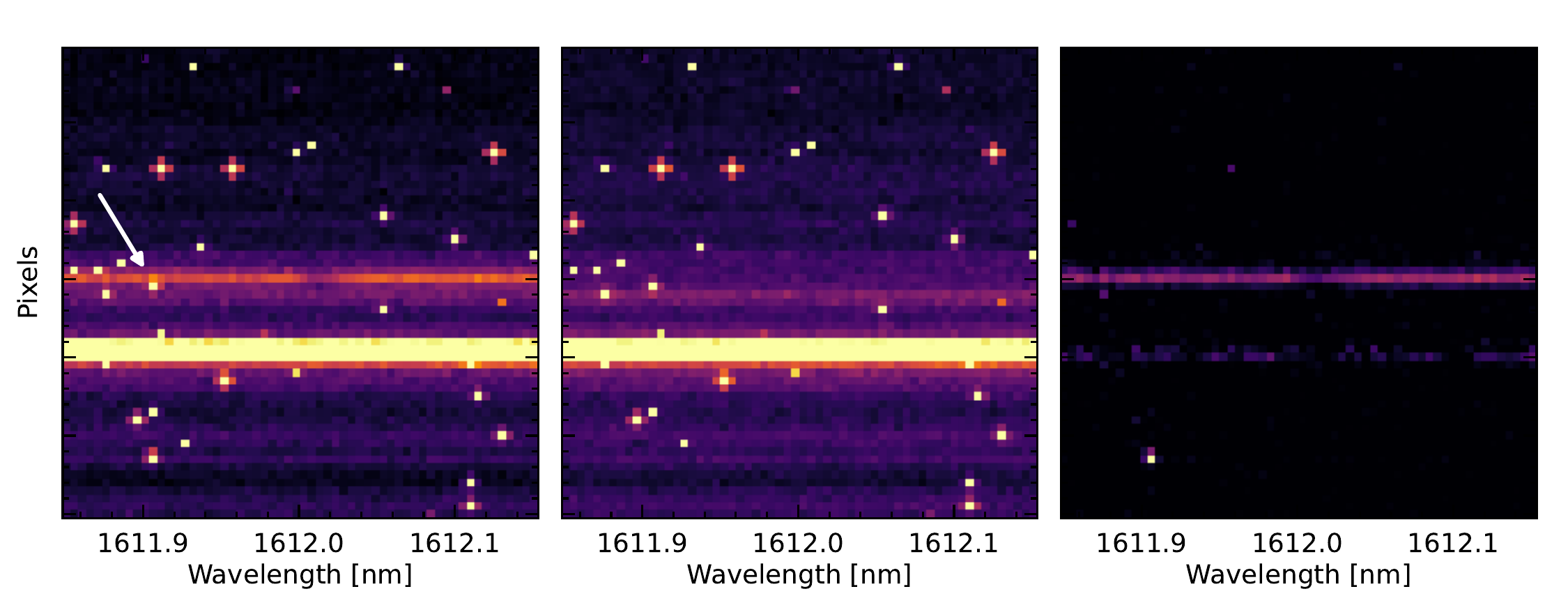}
  \caption{Examples of AO guide fiber leakage into the slit and its impact on science data. Order \#2, detector \#3 at the top, and order \#5, detector \#2 at the bottom. The color scale and normalization are the same for all images, and the bad pixels have not been corrected. In this example, the companion has a visual magnitude of $H = 12.6$ and the images are the sum of two DITs of 600\,s. The position of the companion's signal in the raw data is identified with a white arrow.}
  \label{fig:guide_fiber_leakage}
\end{figure*}

Wavefront aberrations are known to decrease the coupling efficiency of the telescope's PSF in single mode fibers \citep{Jovanovic2015,Otten2021}. During HiRISE commissioning, we estimated the amount of low-order non-common path aberrations (NCPA) at the level of the science fiber by introducing ramps of aberrations using the high-order deformable mirror (HODM) of SPHERE and measuring the flux at the output on the SV camera of \crires. Aberrations were introduced using Zernike modes from Z3 to Z10 (focus, astigmatisms, comas, trefoils, and spherical aberration) with ramps ranging from -100 to +100\,nm\,rms in 15 steps. At each step, an injection optimization was done on the science fiber to ensure proper centering and maximize coupling efficiency. The Zernike modes are projected on 900 of the 990 Karhunen-Loève modes that are used by SAXO to filter spatial frequencies that cannot be applied to the HODM. The modes are then introduced with an offset on the WFS reference slopes. This is the same method used by \citet{Vigan2019} for the compensation of the NCPA with the ZELDA wavefront sensor.

The results of the analysis are presented in Fig.~\ref{fig:ncpa}. In this figure, we plot the normalized flux measured at the output of the fiber as a function of the introduced aberration offset. The curves are all normalized by a common factor, which is the average value for all curves at zero nanometers of aberration. In theory, all curves should cross for a value of zero, but in practice the noise and small differences in the convergence of the centering introduce some variations of the order of 0.4\%. The biggest variation is observed for the astigmatism at 0\degre, which is $\sim$2\% below the other curves. We measure a total of 47\,nm\,rms of optical aberrations at the level of the science fiber. This value is lower than the 70\,nm\,rms reported in Sect.~\ref{sec:fim}, but we include fewer terms than in the Phasics measurement (although aberrations higher than the first order were found to be negligible) and some variations can be expected from the SPHERE instrument itself \citep[see, e.g.,][]{Vigan2022}.

Globally, our analysis shows that the response of the system to static NCPA is relatively flat in the $\pm$40\,nm\,rms range, with losses on the order of 2\% or less. This means that the system is close to an optimal setup in terms of aberrations and that improving the wavefront by compensating the measured NCPA will only provide a limited improvement in the injection efficiency. Of course, in a photon starving regime, any small gain is desirable, but this has to be weighted against the stability of the NCPA and the time required to perform the NCPA calibration. These aspects will be further explored in the future.

\subsection{AO guide fiber leakage}
\label{sec:guide_fiber_leakage}

\begin{figure*}
  \centering 
  \includegraphics[width=1\textwidth]{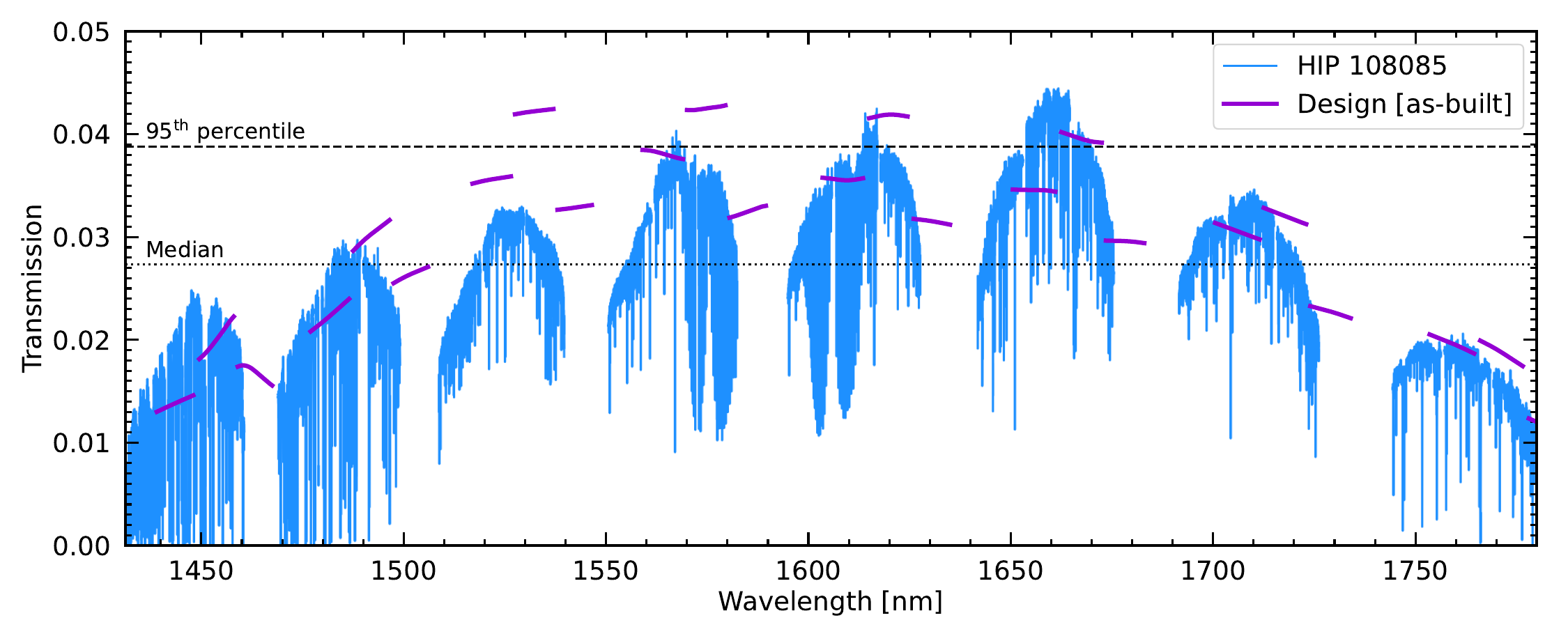}
  \caption{Transmission curve obtained in \texttt{H1567} setting on HIP\,108085 during commissioning in median seeing (seeing of 0.7--0.8\as, unknown $\tau_0$) and clear sky conditions. The design data is based on \citet{Vigan2022spie} with some updates. It corresponds to the as-built design for the final system, i.e. that based on actual transmission measurements for SPHERE, \crires, and most HiRISE optics. The complete transmission budget at a few selected wavelengths is provided in Table~\ref{tab:transmission}.}
  \label{fig:transmission}
\end{figure*}

During the observation of very faint companions as part of commissioning, we discovered the existence of a leakage from the AO guide fiber into the slit, almost at the location of the science fiber. This is illustrated in Fig.~\ref{fig:guide_fiber_leakage} with the observation of a companion that has a visual magnitude of $H = 12.6$, for two different orders and science detectors.

The effect shows a strong variation with wavelength. In the first spectral orders (Fig.~\ref{fig:guide_fiber_leakage}, top), the pattern of the leakage signal is very broad with many spatial features varying in intensity. The signal of the leakage term completely dominates over the signal of the companion, which is strongly attenuated by telluric absorption in this range of wavelengths. In orders closer to the center of the $H$ band (Fig.~\ref{fig:guide_fiber_leakage}, bottom), the leakage signal is more peaked but still presents faint features extending at the location of the science fiber and beyond. The peak flux of the leakage signal is on the order of 1.6--1.8\,ADU/s in both orders, which is at least ten times higher than the signal of the companion.

The origin of the leakage appears relatively obvious when looking at Fig.~\ref{fig:sv_fibers}. In this figure, the diffraction pattern of the AO guide fiber clearly extends in between the ref2 and science fibers. Moreover, the cross-like pattern diffraction indicates that the leakage is strongly amplified by the pupil mask installed on the MACAO deformable mirror. Without this mask, the leakage term would certainly still be present, but probably at a much lower level.

During commissioning, the effect of the leakage was attenuated by acquiring sky backgrounds with the AO guide fiber illuminated. The subtraction of the background is efficient in removing the contribution of the leakage signal (Fig.~\ref{fig:guide_fiber_leakage}, third column), but this cannot remove the induced photon noise. This is visible in the difference images in Fig.~\ref{fig:guide_fiber_leakage}, where a trail of high noise residuals can be identified at the location of the leakage signal's peak. The fact that this leakage term is static, because it was induced by calibration source, could also open the possibility to model and subtract it. But again, this approach will not remove the added photon noise in the data, which will certainly become problematic for companions with higher visual magnitudes.

Different mitigation strategies can be considered. Unfortunately it is not feasible to use one of the reference fibers instead of the science fiber because the reference fibers are located off-axis, or to use one of the dummy fibers because they are not connected to anything. The easiest mitigation strategy is to decrease the intensity of the calibration source that feeds the guide fiber. In the current configuration, an attenuation filter of a factor $\sim$$10^3$ is used in combination with the $H$-band filter in front of the SV camera, and the DIT of the SV camera is set at its minimum value of $\sim$80\,ms. Decreasing the intensity of the source would be compensated by decreasing the attenuation to $\sim$$5 \times 10^2$ or zero (the two other available values) or by increasing the DIT. However, this will have an impact on the target acquisition procedure described in Sect.~\ref{sec:acquisition}, and some careful verification of the updated procedure will be required. A more complex mitigation strategy would be to work in open-loop mode for the duration of individual DITs (typically 600\,s), possibly closing the loop in-between DITs to correct for any temporal or thermal drifts. Another option would be to have a low-pass filter in the halogen source to cut any light above 1.4\,\mic and guide on the SV camera in the $J$ band. The most appropriate mitigation strategy will be investigated in the near future and will be implemented for future observations. Following the implementation of a solution, we will carry out a full assessment of the limitations due to the leakage.

\subsection{Transmission}
\label{sec:transmission}

\begin{table*}
  \caption[]{As-built transmission budget.}
  \label{tab:transmission}
  \centering
  \begin{tabular}{lcccl}
    \hline\hline
    Component           & 1500\,nm & 1600\,nm & 1700\,nm & Description \\
    \hline
    Sky                 & 91\%     & 96\%     & 98\%     & From ESO \texttt{SkyCalc} \\
    Telescope           & 80\%     & 80\%     & 80\%     & 3 reflective surfaces + dust on primary mirror \\
    SPHERE CPI          & 55\%     & 57\%     & 46\%     & SPHERE commissioning measurements \\
    FIM                 & 87\%     & 88\%     & 90\%     & 2 reflective + 4 transmissive optics, as-built AR coatings \\
    Strehl              & 83\%     & 85\%     & 86\%     & Bright guide star ($R = 3$), median conditions, SPHERE NCPA \\
    Pointing error      & 88\%     & 89\%     & 90\%     & 8 mas pointing error \citep{ElMorsy2022} \\
    Coupling efficiency & 76\%     & 76\%     & 76\%     & VLT pupil \\
    Fiber transmission  & 96\%     & 95\%     & 94\%     & Fiber absorption, as-built AR coatings \\
    FEM                 & 91\%     & 92\%     & 93\%     & 1 reflective + 3 transmissive optics, as-built AR coatings \\
    \crires             & 16\%     & 19\%     & 19\%     & AIT laboratory measurements \\
    \hline
    Total               & 2.7\%    & 3.7\%    & 3.1\%    & \\
    \hline
    Measured on sky     & 2.2\%    & 3.3\%    & 3.1\%    & HIP\,108085, 0.7--0.8\as seeing, clear sky conditions \\
    \hline
  \end{tabular}
\end{table*}

The final aspect of performance is the end-to-end transmission of the system. As highlighted by several works in the past \citep[e.g.,][]{Wang2017,Otten2021,Delorme2021}, end-to-end transmission is the main performance driver when combining HCI and HDS.

Transmission measurements were obtained during the last night of commissioning on the bright star HIP\,108085 ($V = 3.01$, $H = 3.43$) in clear sky conditions. The DIMM instrument was not working on that night, but the seeing was estimated to be of the order of 0.7--0.8\as by the VLT/FORS instrument and by the telescope guide probe. Unfortunately, there are no estimations of the coherence time $\tau_0$ for that night. Conditions were very stable during the six 20 second science exposures, followed by equivalent background exposures. The target acquisition and centering on the science fiber was performed using the procedure described in Sect.~\ref{sec:acquisition}, but stopped after the centering of the star. The flux extracted for the star from the \crires data was compared to a black body computed at the temperature of the star ($\Teff \simeq 12\,000$\,K) and normalized using the 2MASS $H$-band filter zero point and the $H$ magnitude of the star. The result of this analysis is presented in Fig.~\ref{fig:transmission}.

We measured a peak transmission for HiRISE of 3.9\%, which was estimated using the 95$^{\mathrm{th}}$ percentile of all the values\footnote{The 95$^{\mathrm{th}}$ percentile is used to avoid the bias induced by telluric absorption in the sprectrum, which is not corrected in these observations.}, and a median value of 2.7\%. These values are perfectly compatible with the expectations from design based on transmission measurements for SPHERE, \crires, and HiRISE optics and fibers \citep{Vigan2022spie}. A breakdown of the transmission budget is provided in Table~\ref{tab:transmission}, and the reader can refer to \citet{Vigan2022spie} for more specific details on the transmission budget.

\subsection{First astrophysical results}
\label{sec:first_result}

To confirm the capability of HiRISE to correctly target and characterize substellar companions, we observed the brown dwarf \object{HD\,984\,B} during the same night as HIP\,108085. \object{HD\,984} is a young (30--200\,Myr) F7 star hosting a brown dwarf \citep{Meshkat2015} with an estimated mass of $61 \pm 4$\,\MJup \citep{Franson2022}. This system was selected based on the brightness of its host star ($H = 6.2$), the moderate contrast of the companion ($\Delta H = 6.4$~mag), and its angular separation of 190\,mas that falls well under the prospect of HiRISE \citep{Vigan2022}.

We acquired six exposures on HD\,984\,B with a DIT of 600\,s each in the \texttt{H1567} spectral setting. Additionally, we also acquired one exposure of 60\,s on the host star and two background exposures of 600\,s. The data are reduced using a custom Python pipeline, which will be described thoroughly in forthcoming publications. Briefly, the pipeline subtracts the background and extracts the companion's flux in each wavelength bin within a fixed aperture. The effect of bad pixels is then removed in the extracted spectrum. The noise is also estimated on the detector in regions devoid of signal. For the wavelength calibration, we use the official \crires pipeline recipe \citep{Dorn2023} without any additional correction. There are, however, some systematic issues with the wavelength calibration that are under investigation and will require additional calibration steps. More generally, the HiRISE pipeline is under active development and will significantly evolve in the near future.

As a preliminary analysis, we simply confirm the detection of HD\,984\,B using a modified cross-correlation function \citep[CCF;][]{Ruffio2019,Wang2021}. In short, we estimate the maximum likelihood value for both the companion and the stellar signal, extracted from the HiRISE data, as a function of RV shift for a given BT-Settl template atmospheric model at the \Teff expected for the companion ($\sim$2\,700\,K, \citealt{Johnson2017}). In this analysis, we used all the orders across the three \crires detectors, covering a wavelength range from 1.43 to 1.78\,\mic.

\begin{figure}
  \centering 
  \includegraphics[width=0.48\textwidth]{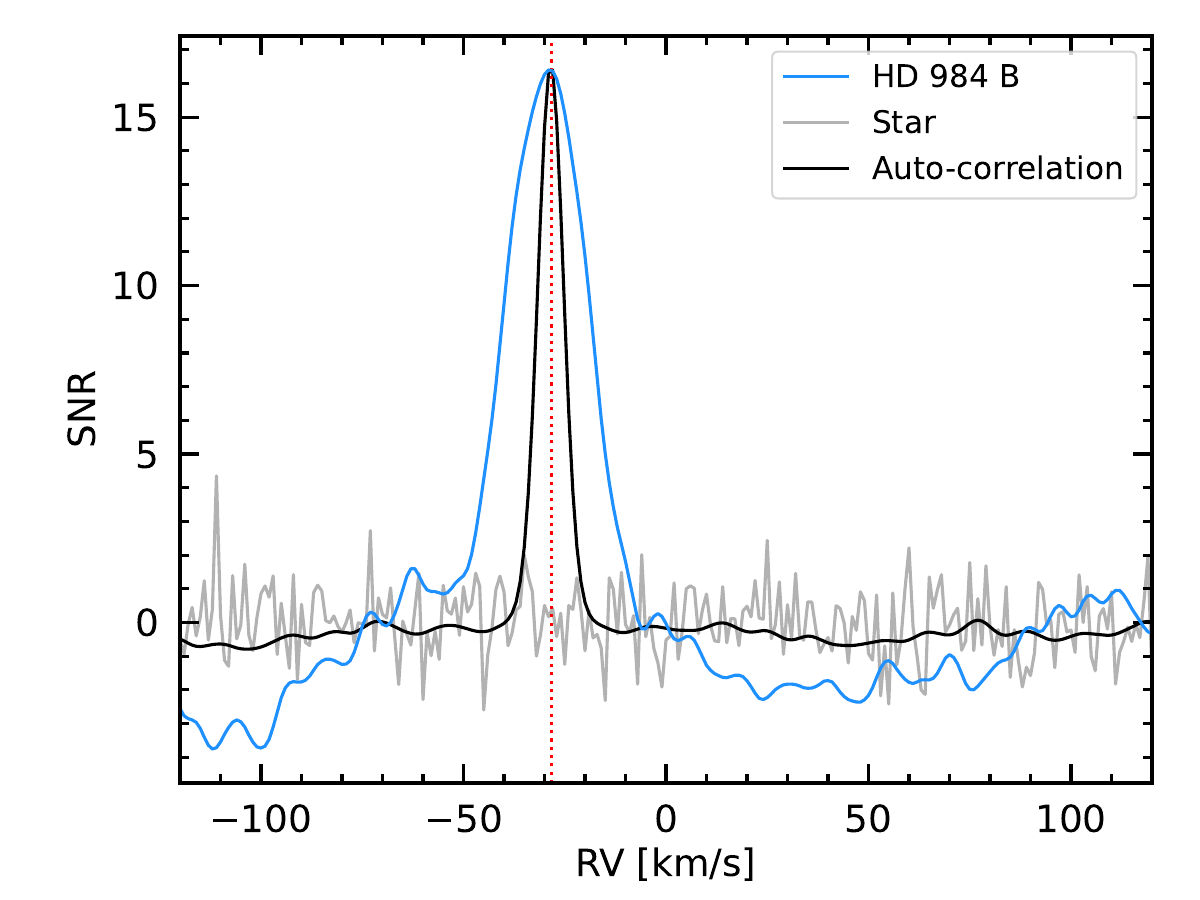}
  \caption{Detection of substellar companion HD\,984\,B using 60\,min of HiRISE commissioning data. The plot shows the normalized CCF between the HiRISE data on the companion and a BT-Settl model at 2\,800\,K in blue, the CCF between the stellar signal and the same model in gray, and the auto-correlation function of the model in black (see Sect.~\ref{sec:first_result} for details). We measure a radial velocity of $-28$\,\kms (red dashed line) and a projected rotational velocity $v\sin i \sim$13\,kms.}
  \label{fig:ccf_HD984B}
\end{figure}

Figure~\ref{fig:ccf_HD984B} presents the results of this modified cross-correlation for the companion and the star. Both CCFs are normalized by their respective standard deviations computed for RV values outside the $\pm$150\,\kms range. We confirm the clear detection of HD\,984\,B at a signal-to-noise ratio (S/N) of $\sim$15 and with an RV shift of $-28$\,\kms, taking into account the radial velocity of the host star and the barycentric corrections. This is in good agreement with the shift expected from orbital predictions based on previous observations (\citealt{Meshkat2015}; \citealt{Johnson2017}; Costes et al. in prep.), which is estimated to $-31$\,\kms using the tool \href{http://whereistheplanet.com}{\texttt{whereistheplanet}} \citep{Wang2021ascl}.

The CCF of the companion appears broadened with respect to the autocorrelation function of the atmospheric template, which is indicative of a significant rotational broadening. We estimate a rotational velocity of $v\sin i \sim$13\,\kms, which is consistent with the rotational velocity measured on $K$-band KPIC data (Costes et al., in prep.).

\section{Conclusions \& prospects}
\label{sec:conclusions}

The HiRISE instrument for the VLT builds on two flagship instruments: the SPHERE exoplanet imager and the \crires high-resolution spectrograph. The goal of this new instrument is the detailed spectral characterization of young EGPs in the $H$ band to better understand their formation, composition, and evolution. 

The system was implemented in a challenging environment, with the major constraint of not impacting the regular operations of SPHERE, \crires, or the telescope. In SPHERE, the FIM has been installed in the NIR arm of the IFS to benefit from the space available there. It includes a tip-tilt mirror to move the image with respect to the science fibers and a tracking camera for monitoring and target acquisition. The FIM is complemented with a dedicated electronics cabinet that contains the instrument workstation, the calibration sources, and the control electronics. Then, an 80 meter fiber bundle is used to link SPHERE and \crires around the UT3. The bundle includes fibers dedicated to science and fibers dedicated to calibrations. All but one fiber are based on the Nufern 1310M-HP telecom fiber, made of fused silica, which offers extremely high transmission in the $H$ band. Finally, the FEM is a simpler module installed in the calibration carrier stage at the entrance of \crires. It is designed to reimage the output of the fiber bundle at the entrance of the instrument with the focal ratio of the VLT. 

We developed an operational model that is a mix between existing observing VLT software templates, some new dedicated templates, and some python procedures. These tools are combined to perform calibrations that are necessary to accurately center the stars' or companions' PSFs on the science fiber, and to maintain this centering over the course of long exposures. One of the operational challenges is related to the MACAO adaptive optics system in \crires that requires a dedicated guide fiber to work in a closed loop and maintain a perfectly flat and stable deformable mirror. The second main challenge is related to the accurate centering of the PSF on the science fiber. For this particular issue, we opted for a procedure that is fully done on the internal calibration source of SPHERE and relies on the DTTS to place the PSF at exactly the same location when switching to the actual star. This procedure has been found to be extremely reliable, but more investigations are required.

Different aspects of calibration and performance have been investigated. In particular, we find that the instrument is not stable when faced with temperature variations, which implies that the PSF will move in front of the science fiber when the temperature varies. The FIM includes a feedback channel that allows the imaging of calibration fibers on top of the science image on the tracking camera. We find that there is a good correlation between the motion of these spots with the image of the science fiber, but there is still a small differential motion that will need to be compensated in the future. This will enable the optimization of the performance during nights where there are significant variations of temperature.

We also measured a total of $\sim$50\,nm\,rms of optical aberrations at the level of the science fiber. These NCPAs could be compensated in the future, depending on stability, but our measurements show that the expected gain is only marginal.

We identified a small leakage signal originating from the AO guide fiber. Currently, the level of this signal is significantly higher than the signal of substellar companions in long exposures. This effect seems to be amplified by the presence of a pupil mask on the deformable mirror of MACAO, which creates diffraction spikes at a location just beneath the science fiber. We will investigate the possibility of decreasing the intensity of the calibration source that feeds the fiber or to work in open loop during science exposures.

Finally, we obtained a first estimation of the end-to-end transmission of the system, which is a driving parameter of the final performance. With a peak transmission of 3.9\%, and a median transmission of 2.7\%, we are almost exactly at the level expected from the design using actual measurements of SPHERE, \crires, and most as-built HiRISE optics. We also presented our first clear astrophysical detection of the brown dwarf companion around HD\,984. These results are extremely encouraging for the execution of an ambitious science program for the understanding of EGPs with HiRISE in the coming years \citep{Otten2021}.

HiRISE on the VLT, KPIC on Keck, and REACH on Subaru, are pathfinder instruments for the future of very-high-contrast imaging science, either on the current telescope \citep[e.g., RISTRETTO on the VLT,][]{Lovis2017} or on future extremely large telescopes \citep[e.g., PCS on the ELT,][]{Kasper2021}. They constitute stepping stones for ground-based instruments that may one day enable the detection of telluric planets around nearby stars.

\begin{acknowledgements}
  The HiRISE team is extremely grateful to ESO and the Paranal observatory staff for their major support throughout the whole project, and especially during the installation and commissioning phase in June/July 2023. Thank you so much everyone! This project has received funding from the European Research Council (ERC) under the European Union's Horizon 2020 research and innovation programme, grant agreements No. 757561 (HiRISE) and 678777 (ICARUS), from the \emph{Commission Spécialisée Astronomie-Astrophysique} (CSAA) of CNRS/INSU, and from the \emph{Action Spécifique Haute Résolution Angulaire} (ASHRA) of CNRS/INSU co-funded by CNES, and from Région Provence-Alpes-Côte d'Azur under grant agreement 2014-0276 (ASOREX).
\end{acknowledgements}

\bibliographystyle{aa}
\bibliography{hirise_first_light}

\end{document}